\documentclass[a4paper,12pt]{article}

\usepackage{comment}
\usepackage[longnamesfirst,sort]{natbib}
\usepackage{fancyvrb}
\usepackage{graphicx}
\usepackage{hyperref}
\usepackage[ragged]{footmisc}
\usepackage{ragged2e}
\usepackage[section]{placeins}
\usepackage{layout}
\usepackage{floatpag}
\usepackage{gensymb}
\usepackage{lscape}
\usepackage{geometry}

\begin{document}

\title{Software engineering and the SP Theory of Intelligence}

\author{J Gerard Wolff\footnote{Dr Gerry Wolff, BA (Cantab), PhD (Wales), CEng, MBCS, MIEEE; CognitionResearch.org, Menai Bridge, UK; \href{mailto:jgw@cognitionresearch.org}{jgw@cognitionresearch.org}; +44 (0) 1248 712962; +44 (0) 7746 290775; {\em Skype}: gerry.wolff; {\em Web}: \href{http://www.cognitionresearch.org}{www.cognitionresearch.org}.}}

\maketitle

\begin{abstract}

This paper describes a novel approach to software engineering derived from the {\em SP Theory of Intelligence} and its realisation in the {\em SP Computer Model}. Despite superficial appearances, it is shown that many of the key ideas in software engineering have counterparts in the structure and workings of the SP system. Potential benefits of this new approach to software engineering include: the automation or semi-automation of software development, with support for programming of the SP system where necessary; allowing programmers to concentrate on `world-oriented' parallelism, without worries about parallelism to speed up processing; support for the long-term goal of programming the SP system via written or spoken natural language; reducing or eliminating the distinction between `design' and `implementation'; reducing or eliminating operations like compiling or interpretation; reducing or eliminating the need for verification of software; reducing the need for validation of software; no formal distinction between program and database; the potential for substantial reductions in the number of types of data file and the number of computer languages; benefits for version control; and reducing technical debt.

\end{abstract}


\section{Introduction}


This paper is about a novel approach to software engineering with potential advantages over standard approaches. It is a considerable revision, expansion and development of preliminary ideas in \citet[Section 6.6]{sp_benefits_apps}. There is an outline description of the SP system in Appendix \ref{outline_of_sp_system_appendix} with pointers to where fuller information may be found.

The novelty of the approach is because it derives from the {\em SP Theory of Intelligence} and its realisation in the {\em SP Computer Model}. Despite superficial appearances, it is shown that many of the key ideas in software engineering have counterparts in the structure and workings of the SP system.

It is envisaged that the SP theory and its realisation in the SP computer model will be the basis for an industrial-strength {\em SP Machine} (Appendix \ref{sp_machine_appendix}) which would be the vehicle for software engineering as described in this paper.

The workings of the SP system may be classified as natural for three main reasons:

\begin{itemize}

    \item {\em Information compression in human learning, perception, and cognition}. The SP system incorporates the principle that much of human learning, perception, and cognition may be understood as information compression. Relevant evidence derives from: research by Fred Attneave \citeyearpar{attneave_1954}, Horace Barlow \citeyearpar{barlow_1959,barlow_1969}, and others, exploring the role of information compression in human perception and cognition. Much additional evidence is described in \citet{sp_compression}.

    \item {\em Information compression in language learning}. A programme of research developing computer models of language learning (summarised in \citet{wolff_1988}) which demonstrates the importance of information compression in learning artificial analogues of natural language.

    \item {\em Modelling aspects of human intelligence}. Although the SP system is little more than the essentially simple concept of SP-multiple-alignment (Appendix \ref{sp_multiple_alignment_appendix}), it has proved to be remarkably versatile in modelling several aspects of human learning, perception, and cognition, as summarised in Appendix \ref{uai_appendix}. In general, the SP system is strongly oriented towards human and thus natural forms of computing.

\end{itemize}

As a preparation for the main body of the paper, the next section relates concepts in ordinary computers to concepts in the SP system. Sections that follow describe several potential advantages of the SP system in software engineering. All sections presuppose some understanding of the structure and workings of the SP system, as outlined in Appendix \ref{outline_of_sp_system_appendix}.

\section{How concepts that are familiar in ordinary computer programming may be seen in the workings of the SP system}\label{relating_conventional_concepts_to_sp_concepts}

Superficially, the workings of an ordinary computer is quite different from the workings of the SP system. Ordinary computers are normally seen to work via the `execution' of `procedures' or `functions' but the SP system works entirely via the compression of information. That the two kinds of processing may be seen to be equivalent is an important insight from the SP programme of research.

This section demonstrates how several of the concepts that are familiar in the programming of ordinary computers may be seen in the workings of the SP system.

\subsection{`Function', `calling of a function', `parameter', and `conditional statement'}\label{program_etc_section}

We begin with a simple example: the kinds of things that need to be done in preparing a meal in a restaurant in response to an order from a customer, excluding any advance preparation of the ingredients.

\subsubsection{An outline of C code for preparing meals in a restaurant}\label{c_code_section}

In the C programming language, relevant functions for preparing meals in a restaurant are shown in outline in Figure \ref{prepare_meal_function_figure}. Here, the highest level structure is the `\texttt{prepare\_meal()}' function at the bottom of the figure, with subordinate functions above it, in accordance with convention.

\begin{figure}[!htbp]
\fontsize{09.00pt}{10.80pt}
\centering
{\bf
\begin{BVerbatim}
void starter(int ST)
{
    if (ST == 0) mussels() ;
    else if (ST == 1) soup() ;
    else avocado() ;
}

void main_course(int MC)
{
    if (MC == 0) lasagna() ;
    else if (MC == 1) beef() ;
    else if (MC == 2) nut-roast() ;
    else if (MC == 3) kipper() ;
    else salad() ;
}

void pudding(int PD)
{
    if (PD == 0) ice_cream() ;
    else if (PD == 1) apple_crumble() ;
    else if (PD == 2) fresh_fruit() ;
    else tiramisu() ;
}

void prepare_meal(int ST, int MC, int PD)
{
    starter(ST) ;
    main_course(MC) ;
    putting(PD) ;
}
\end{BVerbatim}
}
\caption{An outline of C code for preparing a meal in a restaurant.}
\label{prepare_meal_function_figure}
\end{figure}

The top-level function may be called like this: `\texttt{prepare\_meal(0, 4, 1)}'. This has the effect of calling the `\texttt{starter()}' function with the parameter `\texttt{0}', the `\texttt{main\_course(}' function with the parameter `\texttt{4}', and the `\texttt{pudding()}' function with the parameter `\texttt{1}'. As can be seen in the subordinate functions, the parameters have the effect, via conditional statements, of calling the functions `\texttt{mussels()}', `\texttt{salad()}', and `\texttt{apple\_crumble()}'. Each of these may, in an intelligent robot, prepare the corresponding dish, or may at least instruct a person to prepare that dish.

\subsubsection{An outline of an SP grammar for preparing meals in a restaurant}

Figure \ref{prepare_meal_grammar_figure_1} shows an SP {\em grammar}, comprising a collection of SP-patterns, which may be seen as a function for the preparation of meals corresponding to the example in Figure \ref{prepare_meal_function_figure}.

The first SP-pattern in the figure, `\texttt{PM ST \#ST MC \#MC PD \#PD \#PM}', describes the overall structure of the operation of preparing a meal. It is identified by the pair of SP-symbols `\texttt{PM ...~\#PM}' which are mnemonic for ``prepare meal''.

As with the example shown in Figure \ref{prepare_meal_function_figure}, the main steps are the preparation of a starter (`\texttt{ST ...~\#ST}'), the preparation of the main course (`\texttt{MC ...~\#MC}'), and the preparation of a pudding (`\texttt{PD ...~\#PD}'). Corresponding SP-patterns are shown in the second and subsequent rows in the figure.

\begin{figure}[!htbp]
\fontsize{09.00pt}{10.80pt}
\centering
{\bf
\begin{BVerbatim}
PM ST #ST MC #MC PD #PD #PM  | Prepare meal
ST 0 mussels #ST             | Starter: prepare a dish of mussels
ST 1 soup #ST                | Starter: prepare a bowl of soup
ST 2 avocado #ST             | Starter: prepare an avocado dish
MC 0 lasagna #MC             | Main course: prepare a lasagna dish
MC 1 beef #MC                | Main course: prepare a beef dish
MC 2 nut-roast #MC           | Main course: prepare a nut-roast dish
MC 3 kipper #MC              | Main course: prepare a kipper
MC 4 salad #MC               | Main course: prepare a salad
PD 0 ice cream #PD           | Pudding: prepare ice cream
PD 1 apple_crumble #PD       | Pudding: prepare apple crumble
PD 2 fresh_fruit #PD         | Pudding: prepare fresh fruit
PD 3 tiramisu #PD            | Pudding: prepare tiramisu
\end{BVerbatim}
}
\caption{An SP grammar comprising a set of SP-patterns representing in outline the kinds of procedures involved in preparing a meal for a customer in a restaurant. To the right of each SP-pattern is an explanatory comment, beginning with `\texttt{|}', which is not part of the SP-pattern.}
\label{prepare_meal_grammar_figure_1}
\end{figure}

\subsubsection{Building SP-multiple-alignments via information compression}

\begin{figure}[!htbp]
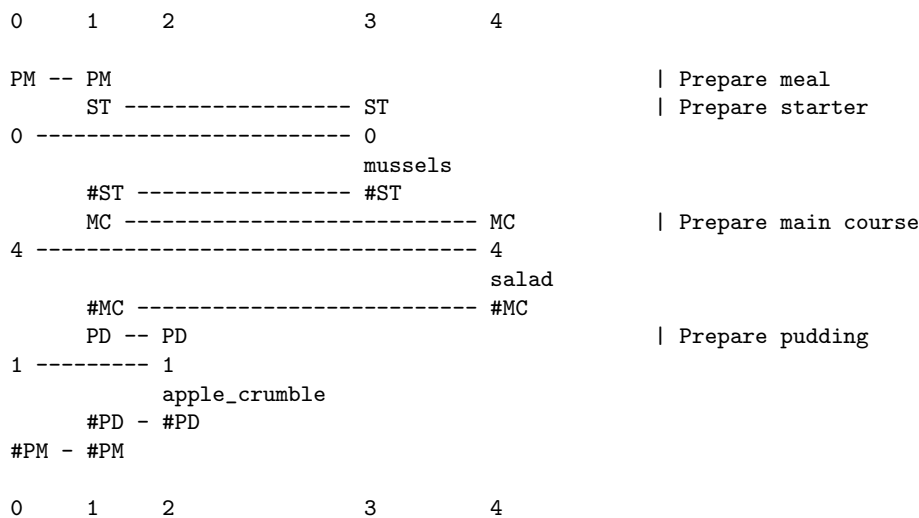

\fontsize{09.00pt}{10.80pt}
\centering
{\bf
\begin{BVerbatim}
0     1     2               3         4

PM -- PM                                           | Prepare meal
      ST ------------------ ST                     | Prepare starter
0 ------------------------- 0
                            mussels
      #ST ----------------- #ST
      MC ---------------------------- MC           | Prepare main course
4 ----------------------------------- 4
                                      salad
      #MC --------------------------- #MC
      PD -- PD                                     | Prepare pudding
1 --------- 1
            apple_crumble
      #PD - #PD
#PM - #PM

0     1     2               3         4
\end{BVerbatim}
}
\caption{The best SP-multiple-alignment created by the SP Computer Model with the New SP-pattern, `\texttt{PM 0 4 1 \#PM}', and the set of Old SP-patterns shown in Figure \ref{prepare_meal_grammar_figure_1}.}
\label{prepare_meal_ma_figure_1}
\end{figure}

To see how this grammar functions in practice, consider the SP-multiple-alignment shown in Figure \ref{prepare_meal_ma_figure_1}.\footnote{Just to confuse matters, this SP-multiple-alignment has been rotated by $90\degree$ compared with the SP-multiple-alignment shown in Figure \ref{parsing_figure}. These two versions of an SP-multiple-alignment are entirely equivalent. The choice between them depends entirely on what fits best on the page.}

This SP-multiple-alignment is the best one created by the SP Computer Model with the New SP-pattern, `\texttt{PM 0 4 1 \#PM}', processed in conjunction with Old SP-patterns shown in Figure \ref{prepare_meal_grammar_figure_1}. Here, the New SP-pattern may be seen as an economical description of what the customer ordered: a starter comprising a dish of mussels, represented by the code `\texttt{0}'; a main course chosen to be a salad, represented by the code `\texttt{4}'; and a pudding which in this case is apple crumble, represented by the code `\texttt{1}'.

Assuming that each of the SP-symbols `\texttt{mussels}', `\texttt{salad}', and `\texttt{apple\_crumble}', represents the execution of instructions for preparing the corresponding dish, or is at least an instruction to a person to prepare that dish, the whole SP-multiple-alignment may be seen to achieve the effect of preparing what the customer has ordered, much as with the example discussed in Section \ref{c_code_section}.

\subsubsection{How the concepts of `function', `calling of a function', `parameter', and `conditional statement' may be seen in the workings of the SP system}

This example shows how the concepts `function', `calling of a function', `parameter', and `conditional statement' may be seen in the workings of the SP system:

\begin{itemize}

    \item As mentioned earlier, the whole grammar in Figure \ref{prepare_meal_grammar_figure_1} may be seen as a {\em function} for preparing a meal to meet a given order, like the outline code shown in Figure \ref{prepare_meal_function_figure}.

    \item Of the remaining SP-patterns in Figure \ref{prepare_meal_grammar_figure_1}, each group of SP-patterns that begin with the same SP-symbol, such as `\texttt{ST 0 mussels \#ST}', `\texttt{ST 1 soup \#ST}', and `\texttt{ST 2 avocado \#ST}', may be seen as a subordinate {\em function} that is {\em called} from the higher-level function `\texttt{PM ST \#ST MC \#MC PD \#PD \#PM}'.

        Although the example does not illustrate the point, it should be clear that each subordinate function may itself call one or more lower-level functions, and so on through as many levels as may be required. As we shall see in Section \ref{recursion_section}, recursion is also possible.

    \item Each of the code SP-symbols `\texttt{0}', `\texttt{4}', and `\texttt{1}', in the New SP-pattern `\texttt{PM 0 4 1 \#PM}', may be seen as a {\em parameter} to the top-level function.

    \item Since the code SP-symbol `\texttt{0}' has the effect of selecting the SP-pattern `\texttt{ST 0 mussels \#ST}' from the set of SP-patterns `\texttt{ST 0 mussels \#ST}', `\texttt{ST 1 soup \#ST}', and `\texttt{ST 2 avocado \#ST}', the process of selection may be seen to achieve the effect of a {\em conditional statement} or {\em if-then rule} in an ordinary computer program, something like the C statement `\texttt{if (ST == 0) mussels() ;}' in Figure \ref{prepare_meal_function_figure}, meaning ``If the value of `\texttt{ST}' is `0', perform the subordinate function `\texttt{mussels()}', which itself means ``prepare a dish of mussels''. Much the same may be said, {\em mutatis mutandis}, about the code SP-symbols `\texttt{4}', and `\texttt{1}'.

\end{itemize}

\subsection{Variables, values, and types}\label{variables_values_types_section}

In addition to the programming concepts already considered, the concepts `variable', `value', and `type' may be seen in the workings of the SP system.

Consider, for example, the SP grammar shown in Figure \ref{salad_grammar_figure}. This is an expansion of the ``salad'' main course entry in the grammar shown in Figure \ref{prepare_meal_grammar_figure_1}. Instead of simply giving the name of the dish, this grammar provides for choices of ingredients in four categories: salad leaves (`\texttt{L ...~\#L}'), root vegetables (`\texttt{R ...~\#R}'), garnish (`\texttt{G ...~\#G}'), and dressing (`\texttt{D ...~\#D}').

\begin{figure}[!htbp]
\fontsize{09.00pt}{10.80pt}
\centering
{\bf
\begin{BVerbatim}
MC salad L #L R #R G #G D #D #MC
L 0 lettuce #L
L 1 beetroot_leaves #L
L 2 water_cress #L
L 3 spinach #L
R 0 carrot #R
R 1 potato #R
R 2 parsnip #R
G 0 nuts #G
G 1 saltanas #G
D 0 mayonnaise #D
D 1 vinaigrette #D
D 2 thousand_island #D
\end{BVerbatim}
}
\caption{An SP grammar comprising a set of SP SP-patterns representing in simplified form the kinds of things that may go in a salad. {\em Key:} MC = main course; L = salad leaves; R = root vegetables; G = garnish; D = dressing.}
\label{salad_grammar_figure}
\end{figure}

When the SP Computer Model is run with the New SP-pattern `\texttt{MC 2 1 0 1 \#MC}' and Old SP-patterns comprising the SP-patterns shown in Figure \ref{salad_grammar_figure}, the best SP-multiple-alignment created by the SP Computer Model is the one shown in Figure \ref{salad_ma_figure}.

\begin{figure}[!htbp]
\fontsize{09.00pt}{10.80pt}
\centering
{\bf
\begin{BVerbatim}
0     1       2             3        4      5

MC -- MC
      salad
      L ----- L
2 ----------- 2
              water_cress
      #L ---- #L
      R ------------------- R
1 ------------------------- 1
                            potato
      #R ------------------ #R
      G ---------------------------- G
0 ---------------------------------- 0
                                     nuts
      #G --------------------------- #G
      D ----------------------------------- D
1 ----------------------------------------- 1
                                            vinaigrette
      #D ---------------------------------- #D
#MC - #MC

0     1       2             3        4      5
\end{BVerbatim}
}
\caption{The best SP-multiple-alignment created by the SP Computer Model with the New SP-pattern `\texttt{MC 2 1 0 1 \#MC}' and the set of Old SP-patterns shown in Figure \ref{salad_grammar_figure}.}
\label{salad_ma_figure}
\end{figure}

In this example:

\begin{itemize}

    \item Within the SP-pattern `\texttt{MC salad L \#L R \#R G \#G D \#D \#MC}' (column 1), each of the pair of SP-symbols `\texttt{L \#L}', `\texttt{R \#R}', `\texttt{G \#G}', and `\texttt{D \#D}', may be seen to represent the concept {\em variable} because they are slots where values may be inserted.

    \item The effect of the SP-multiple-alignment is to assign `\texttt{water\_cress}' to the first slot, `\texttt{potato}' to the second slot, `\texttt{nuts}' to the third slot, and `\texttt{vinaigrette}' to the fourth slot. Those four things may be seen to be {\em values}, each one assigned to an appropriate variable.

    \item For each of the four variables, its {\em type}---meaning the range of values that it may take---may be seen to be defined by the grammar shown in Figure \ref{salad_grammar_figure}. For example, possible values for the variable `\texttt{L \#L}' may be seen to be `\texttt{lettuce}', `\texttt{beetroot\_leaves}', `\texttt{water\_cress}', and `\texttt{spinach}'. Likewise for the other three variables.

\end{itemize}

\subsection{Structured programming}\label{structured_programming_section}

An established feature of software engineering today, which is now partly but not entirely subsumed by object-oriented programming (next), is `structured programming' \citep{jackson_1975} in which the central idea is that programs should comprise well-defined structures which should reflect the structure of the data that is to be processed and should {\em never} use the `goto' statement of an earlier era.

To a large extent, the SP system incorporates the principles of structured programming, since unsupervised learning in the SP system creates structures that reflect the structure of incoming data. And other kinds of processing in the SP system, such as pattern recognition or reasoning, is achieved by recognising and processing similar structures in new data, without the use of anything like a `goto' statement.

\subsection{Object-oriented design or programming}\label{oo_design_programming_section}

From its introduction in the {\em Simula} computer language \citep{birtwistle_etal_1973}, `object-oriented programming' (OOP) and the closely-related `object-oriented design' (OOD) have become central in software engineering and in such widely-used programming languages as C++ and Java.

Key ideas in OOP/OOD are that the structure of each computer program should reflect the objects to which it relates---people, packages, fork-lift trucks, and so on---and the classes and subclasses in which each object belongs. This not only helps to make computer programs easy to understand but it means that the features of any specific object may be `inherited' from the classes and subclasses to which it belongs.

Inheritance applies to all the objects in a given class, meaning that there is an overall saving or compression of information compared with what would be needed without inheritance. In this respect, OOP/OOD is very much in keeping with the central importance of information compression in the SP system.

\begin{figure}[!htbp]
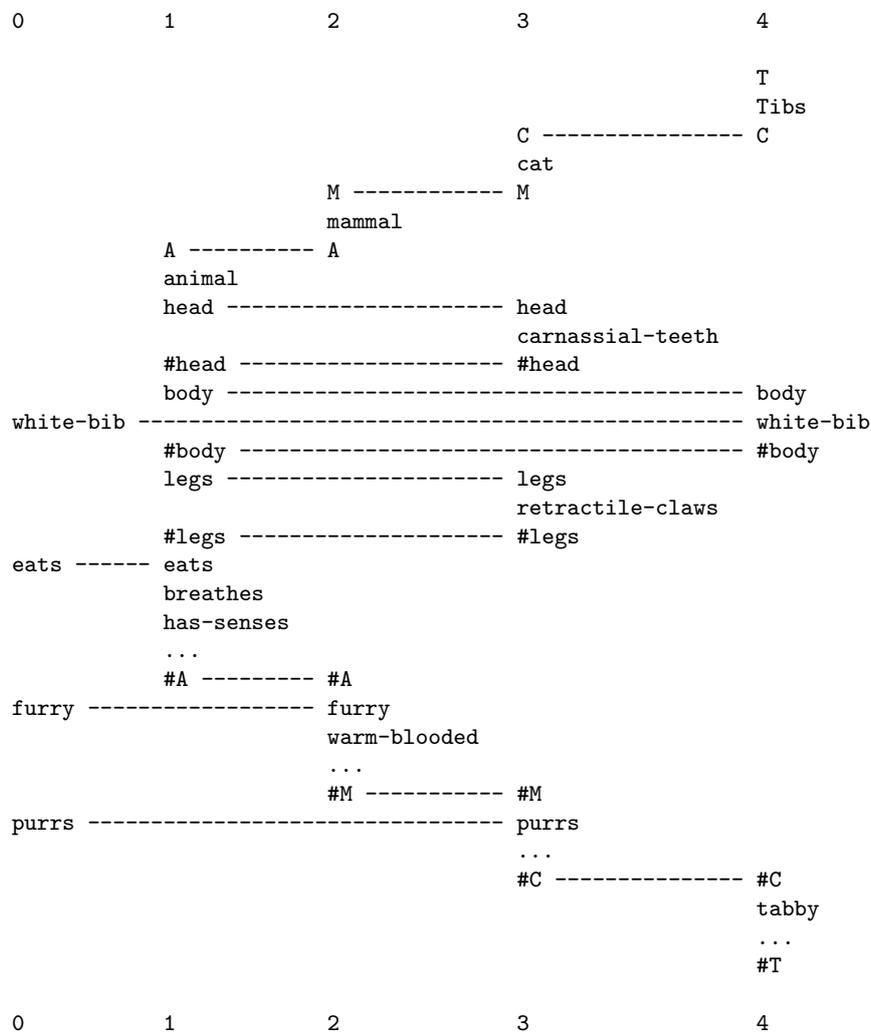

\fontsize{09.00pt}{10.80pt}
\centering
{\bf
\begin{BVerbatim}
0           1            2              3                  4

                                                           T
                                                           Tibs
                                        C ---------------- C
                                        cat
                         M ------------ M
                         mammal
            A ---------- A
            animal
            head ---------------------- head
                                        carnassial-teeth
            #head --------------------- #head
            body ----------------------------------------- body
white-bib ------------------------------------------------ white-bib
            #body ---------------------------------------- #body
            legs ---------------------- legs
                                        retractile-claws
            #legs --------------------- #legs
eats ------ eats
            breathes
            has-senses
            ...
            #A --------- #A
furry ------------------ furry
                         warm-blooded
                         ...
                         #M ----------- #M
purrs --------------------------------- purrs
                                        ...
                                        #C --------------- #C
                                                           tabby
                                                           ...
                                                           #T

0           1            2              3                  4
\end{BVerbatim}
}
\caption{The best SP-multiple-alignment found by the SP Computer Model, with the New SP-pattern `\texttt{white-bib eats furry purrs}' shown in column 1, and a set of Old SP-patterns representing different categories of animal and their attributes shown in columns 1 to 4. Reproduced with permission from Figure 15 in \protect\citet{sp_extended_overview}.}
\label{class_hierarchy_figure}
\end{figure}

In Figure \ref{class_hierarchy_figure}, the SP-multiple-alignment produced by the SP Computer Model, with the New SP-pattern `\texttt{white-bib eats furry purrs}' and a set of Old SP-patterns representing different categories of animal and their attributes, shows how a previously-unknown entity with features shown in the New SP-pattern in column 1 may be recognised at several levels of abstraction: as an animal (column 1), as a mammal (column 2), as a cat (column 3) and as the specific cat ``Tibs'' (column 4). These are the kinds of classes used in ordinary systems for OOP/OOD.

From this SP-multiple-alignment, we can see how the entity that has now been recognised {\em inherits} unseen characteristics from each of the levels in the class hierarchy: as an animal (column 1) the creature `\texttt{breathes}' and `\texttt{has-senses}', as a mammal it is `\texttt{warm-blooded}', as a cat it has `\texttt{carnassial-teeth}' and `\texttt{retractile-claws}', and as the individual cat Tibs it has a `\texttt{white-bib}' and is `\texttt{tabby}'.

\subsection{Recursion}\label{recursion_section}

The SP system does not provide for the repetition of procedures via these kinds of statement: {\em while ...}, {\em do ...~while ...}, {\em for ...}, or {\em repeat ...~until ...}. But the same effect may be achieved via recursion, as illustrated in Figure \ref{recursion_ma_figure}.

\begin{figure}[!htbp]
\fontsize{06.00pt}{07.20pt}
\centering
{\bf
\begin{BVerbatim}
0     1             2     3             4             5     6             7     8             9     10

pg ---------------- pg
                    a ----------------------------------------------------------------------------- a
a6 ------------------------------------------------------------------------------------------------ a6
                                                                                                    procedure_A
                    #a ---------------------------------------------------------------------------- #a
                    ri ------------------------------ ri
                                                      ri1
                                                      ri ---------------- ri
                                                                          ri1
                                                                          ri ---------------- ri
                                                                                              ri1
                                                                                              ri
                                                                                              #ri
                                                                                b ----------- b
b1 ---------------------------------------------------------------------------- b1
                                                                                procedure_B
                                                                                #b ---------- #b
                                                                          #ri --------------- #ri
                                                            b ----------- b
b1 -------------------------------------------------------- b1
                                                            procedure_B
                                                            #b ---------- #b
                                                      #ri --------------- #ri
                                        b ----------- b
b1 ------------------------------------ b1
                                        procedure_B
                                        #b ---------- #b
                    #ri ----------------------------- #ri
      c ----------- c
c4 -- c4
      procedure_C
      #c ---------- #c
                    d --- d
d3 ---------------------- d3
                          procedure_D
                    #d -- #d
#pg --------------- #pg

0     1             2     3             4             5     6             7     8             9     10
\end{BVerbatim}
}
\caption{An SP-multiple-alignment produced by the SP Computer Model showing how, via recursion mediated by a self-referential SP-pattern (which is `\texttt{ri ri1 ri \#ri b \#b \#ri}' in this example), the SP system may model the repetition of a procedure or function, which in this examples is shown as `\texttt{procedure\_B}'.}
\label{recursion_ma_figure}
\end{figure}

In the figure, the SP-symbols `\texttt{a6 b1 b1 b1 c4 d3}' in the New SP-pattern in row 0 (`\texttt{pg a6 b1 b1 b1 c4 d3 \#pg}') may be seen as parameters for the SP `program' or grammar for this example (not shown on this occasion).

One point of interest here is that the SP-pattern `\texttt{ri ri1 ri \#ri b \#b \#ri}' (which appears in columns 5, 7, and 9) is recursive because it is self-referential---because the pair of SP-symbols `\texttt{ri \#ri}' within the larger SP-pattern `\texttt{ri ri1 ri \#ri b \#b \#ri}' may be matched and unified with the same two SP-symbols at the beginning and end of that larger SP-pattern. Hence, the larger SP-pattern contains a reference to itself.

Another point of interest is that the recursive SP-pattern `\texttt{ri ri1 ri \#ri b \#b \#ri}', and its connected SP-pattern `\texttt{b b1 procedure\_B \#b}', each occur 3 times in the SP-multiple-alignment in Figure \ref{recursion_ma_figure}, although each of them only occurs once in the grammar for this SP-multiple-alignment. Hence, the grammar is relatively compressed compared with what would be needed if all possible repetitions were stored explicitly.

With any kind of recursion, something is needed to tell the system when to stop the repetition. In our example, the number of repetitions is specified explicitly by the three instances of the SP-symbol `\texttt{b1}' within the New SP-pattern. Other devices may also be used.

\section{The full or partial automation of software development}\label{full_partial_automation_se_section}

This and the following main sections describe potential advantages of the SP system in software engineering compared with software engineering with conventional computers.

This section considers the full or partial automation of software and the associated issue of generalisation in software, and how to avoid under- and over-generalisaton.

\subsection{Automation of software development}

Assuming that the SP Machine (Appendix \ref{sp_machine_appendix}) has been developed to the stage where it has robust abilities for unsupervised learning with both one-dimensional and two-dimensional SP-patterns, and assuming that residual problems in that area have been solved (Appendix \ref{unsupervised_learning_appendix}), the SP Machine is likely to prove useful in both the automatic and semi-automatic creation of software, discussed in this and the following subsection.

At least two things suggest that such possibilities are credible:

\begin{itemize}

    \item As noted in Section \ref{structured_programming_section} and Appendix \ref{sp_cc_differences_in_rk_appendix}, it has been recognised for some time, in connection with the concept of ``structured programming'', that the structure of software should mirror the structure of the data that it is designed to process \citep{jackson_1975}. This fits well with the observation that in forms of unsupervised learning such as grammatical inference, the structure of the resulting grammar reflects the structure of the data from which it was derived.

    \item In the same vein, in connection with ``object-oriented design'' and ``object-oriented programming'' (Section \ref{oo_design_programming_section}), it is well-established that a well-structured program should reflect the structure of entities and classes of entities that are significant in the workings of the program. This fits well with the observation that unsupervised learning in the SP system appears to conform to the `DONSIC' principle \citep[Section 5.2]{sp_extended_overview}: {\em the discovery of natural structures via information compression}---where `natural' means aspects of our environment such as `objects' which we perceive to be natural.

\end{itemize}

\subsubsection{Example: learning in an autonomous robot}\label{learning_in_autonomous_robot_section}

Perhaps the best example of how the SP system may facilitate automatic programming is in autonomous robots that learn continually via their senses, much as people do \citep{sp_autonomous_robots}. Here, the robot's ever-increasing store of knowledge, together with any in-built motivations, provide the basis for many potential inferences (\citet[Chapter 7]{wolff_2006}, \citet[Section 10]{sp_extended_overview}) and, perhaps more important in the present context, the creation of one or more plans (\citet[Chapter 8]{wolff_2006}, \citet[Section 12]{sp_extended_overview}), each one of which may be regarded as a program to guide the robot's actions.

The potential for this kind of development raises important issues about how much autonomy should be granted to any robot and how external controls may be applied. Pending the resolution of such issues, there is potential in the SP system for more humdrum kinds of automatic programming, as described in the next two subsections.

\subsubsection{Example: processing data received by the SKA}\label{ska_example_section}

The fully automatic creation of software should be possible in situations where there is a body of data that represents the entire problem or a realistically large sample of it. An example is the large volumes of data that will be gathered by the Square Kilometre Array (SKA)\footnote{See, for example, ``Square Kilometre Array'', {\em Wikipedia}, \href{http://bit.ly/2t16xxW}{bit.ly/2t16xxW}, retrieved 2017-07-15.} when it is completed.

With data like this, unsupervised learning by the SP Machine should build grammars that represent entities and classes of entity---such as stars and galaxies---in two dimensions at least, and possibly in three dimensions. And its grammars should also embrace `procedural' or `process' regularities in the time dimension.

Any such grammar may be seen as a `program' for the analysis of similar kinds of data in the future. A neat feature of the SP system is that the SP-multiple-alignment construct serves not only in unsupervised learning but also, without modification, in such operations as SP-pattern recognition, reasoning, and more (Appendix \ref{versatility_in_intelligence_appendix}).

With an area of application like the processing of data received by the SKA, it may of course happen that significant structures or events---such as supernovas or gamma-ray bursts---do not appear in any one sample of data. For that reason, unsupervised learning should be an ongoing process, much as in people, so that the system may gain progressively more knowledge of its target environment as time goes by.

Some more observations relating to this example are described in Section \ref{possible_augmentations_section}.

\subsubsection{Example: programming by demonstration}\label{programming_by_demonstration_example}

Another situation where the SP Machine may achieve fully-automatic creation of software is with a technique for programming robots called ``programming by demonstration''.\footnote{See, for example, ``Programming by demonstration'', {\em Wikipedia}, \href{http://bit.ly/2v3phy8}{bit.ly/2v3phy8}, retrieved 2017-07-15.}

As an example, a person who is skilled at some operation in the building of a car (such as paint-spraying the front of the car) may take the `hand' of a robot and guide it through the sequence of actions needed to complete the given operation. Here, signals from sensors in various parts of the robot's arm, including the robot's actuators or `muscles', would be recorded and the record would constitute a preliminary kind of `program' of the several positions of the arm and actuators that are needed to complete the operation.

Any such preliminary program may be processed by the SP system to convert it into something that more closely resembles an ordinary program, with the equivalent of subroutines, repetition of operations, and conditional statements. To allow for acceptable variations in the task, there should also be appropriate generalisation from the raw data, as described in Section \ref{generalisation_section}.

Some more observations relating to this example are described next.

\subsubsection{Possible augmentations}\label{possible_augmentations_section}

An assumption behind the two examples just described is that the grammar or program created via unsupervised learning would do everything that is needed.

In many cases, this would probably be true. This is because of a neat feature of the SP system: that the SP-multiple-alignment subsystem is not only an important part of unsupervised learning but is also the key to such operations as pattern recognition, several kinds of reasoning, retrieval of information, and problem solving (Appendix \ref{versatility_in_intelligence_appendix}). With the SKA example (Section \ref{ska_example_section}, these kinds of operations may be all that is required. With the programming-by-demonstration example (Section \ref{programming_by_demonstration_example}), the program created via unsupervised learning may function directly in controlling the robot arm.

But the user of the SKA system might want to do such things as showing stars in red, galaxies in green, and so on. And the user of the programming-by-demonstration system might want to add some bells and whistles such as playing musical sounds as the robot works.

Clearly, such augmentations fall outside what could be created automatically via unsupervised learning. They take us into to the realm of semi-automatic creation of software, discussed next.

\subsection{Semi-automatic creation of software}\label{semi-full_partial_automation_se_section}

With some kinds of application, it seems unlikely that the creation of relevant software could be fully automated in the foreseeable future. One example is the kinds of augmentation to an automatically-created program described in Section \ref{possible_augmentations_section}. Another example is the kind of software that is needed to manage a business---with knowledge of people, vehicles, furniture, packages, warehouses, relevant rules and regulations, and so on.

With the latter kind of problem, there appears to be potential for the system to assist in the refinement of human-created software by detecting redundancies in any draft design, and inconsistencies from one part of the design to another. On the assumption that the software is developed using SP-patterns and is hosted on an SP Machine (as outlined in Section \ref{non-automatic_programming_software_section}), then the SP Machine may be a vehicle for verification and validation of the software as described in Sections \ref{verification_section} and \ref{validation_section}.

At some point in the future, it is conceivable that knowledge about how a business operates may, at some stage, be built up by an intelligent autonomous robot of the kind described in \citet{sp_autonomous_robots} that is allowed to explore different areas of the business, observing the kinds entity and operation that are involved, asking questions, and so on. But for the foreseeable future, it seems likely that any software that may have been created by such a robot would need to be augmented and refined by people.

\subsection{Generalisation and the avoidance of under- and over-generalisation}\label{generalisation_section}

As a rule, any given computer program is more general than any set of examples that it may process. For example, an ordinary spreadsheet program can work with millions or perhaps billions of different sets of data, far more than it would ever be used for in practice.

Since we have been considering the possibility that software may be created automatically or semi-automatically in the manner of unsupervised learning (Sections \ref{full_partial_automation_se_section} and \ref{semi-full_partial_automation_se_section}), we need to consider how the system would generalise correctly from the examples it has been given, without either under-generalisation (sometimes called `overfitting') or over-generalisation (sometimes called `underfitting').

The SP system provides an answer outlined in \citet[Section 5.3]{sp_extended_overview},\footnote{That accountcl only mentions over-generalisation but it appears that the same procedure will apply to the avoidance of under-generalisation.} with some supporting evidence. In brief, it appears that correct generalisation may be achieved, without either under- or over-generalisation, like this:

\begin{enumerate}

    \item Given a body of raw data, {\bf I}, compress it as much as possible with the program for unsupervised learning.

    \item Divide the resulting compressed version of {\bf I} into two parts: a {\em grammar}, {\bf G}, which represents the recurring features of {\bf I}, and an {\em encoding}, {\bf E}, of {\bf I} in terms of {\bf G}.

    \item Discard {\bf E} and retain {\bf G}.

\end{enumerate}

Here, {\bf G} may be seen to be a program for processing {\bf I} and for processing many other bodies of data with the same general characteristics as {\bf I}, without either under- or over-generalisation.

\section{Non-automatic programming of the SP system}\label{non-automatic_programming_software_section}

If or when the automatic creation of software is not feasible, or if something more than small revisions are needed with software that has been created semi-automatically, then something like ordinary programming will be needed.

In principle, this can be done using SP-patterns directly. But, mainly for reasons of human psychology, some kind of `syntactic sugar' or other aids may be helpful for programmers. Here are four possibilities:

\begin{itemize}

    \item With an SP-pattern like `\texttt{NP D \#D N \#N \#NP}' in row 4 of Figure \ref{parsing_figure}, it may he helpful if, when the first SP-symbol (`\texttt{NP}') has been typed in, the programming environment would automatically insert the balancing last SP-symbol (`\texttt{\#NP}').

    \item Unless or until programmers become used to how things are done in the SP system, it may be helpful to create a programming environment in which SP concepts are presented in a manner that resembles ordinary programming concepts, as described in Section \ref{relating_conventional_concepts_to_sp_concepts}.

    \item Instances of the object-oriented concept of a class-inclusion hierarchy (Section \ref{oo_design_programming_section}), and instances of any part-whole hierarchy (dividing an object into its parts and subparts) may be represented graphically and implemented with equivalent sets of SP-patterns.

\end{itemize}

There will also be a need for programmers to specify aspects of parallel and sequential processing, as described in Section \ref{sp_advantage_with_parallel_processing_section}, next.

\section{A potential advantage of the SP Machine in the application of parallel processing}\label{sp_advantage_with_parallel_processing_section}

In the application of parallel processing in the SP Machine, it is important to distinguish between two kinds of parallelism:

\begin{itemize}

    \item {\em World-oriented parallelism}. World-oriented parallelism means the kind of parallelism that one might observe in the world, including the activities of people: it may rain at the same time as the wind blows; the players in a game of football are all doing different things at the same time; a cook may prepare the icing for a cake at the same time as the cake is baking; and so on.

    \item {\em Machine-oriented parallelism}. In the workings of a computer, there may be parallel processing in the MapReduce model, in pipelining, in SIMD parallelism, in MIMD parallelism, and so on.

\end{itemize}

In the programming of an ordinary high-parallel supercomputer or high-parallel computing cluster, both kinds of parallelism may be applied, with little or no distinction between them. For example, a programmer who is developing a flight simulator may adopt machine-oriented parallelism for such processing as matrix multiplication and apply world-oriented parallelism in modelling the many processes that are involved, largely in parallel, in a plane's flight.

A potential advantage of the SP machine is that programmers of such a machine may be largely relieved of the need to worry about machine-oriented parallelism and may concentrate on the programming of world-oriented parallelims.

This potential advantage arises because of evidence that the SP system has potential to serve as a {\em universal framework for the representation and processing of diverse kinds of knowledge} (UFK) \cite[Section III]{sp_big_data}.

The evidence is that, already, one relatively simple conceptual and computational framework---SP-multiple-alignment---demonstrates versatility in aspects of intelligence, versatility in the representation of diverse forms of knowledge, with clear potential for the seamless integration of diverse aspects of intelligence and diverse forms of knowledge, in any combination (see \cite[Sections 4, 5, and 6]{sp_intro_2018} and pointers from there). And there are reasons to believe that that versatility may be extended.

If or when the SP system can be developed with full human-like versatility, taking full advantage of machine-oriented parallel processing, then it seems likely that programmers can largely relieved of concerns about that kind of parallelism and, for any given area of application, they may concentrate world-oriented parallelism in that domain.

\section{Programming via natural language}\label{programming_via_nl_section}

One of the strengths of the SP system is in the processing of natural language, mentioned in Appendices \ref{versatility_in_rk_appendix} and \ref{versatility_in_intelligence_appendix}, and described in more detail in \citet[Section 8]{sp_extended_overview} and \citet[Chapter 5]{wolff_2006}.

There is clear potential in the SP system for developing human-level processing of writing, and ultimately speech, but there will be some difficult hurdles to overcome, probably requiring a two-pronged attack: working on problems in the processing of natural language together with problems in the unsupervised learning of syntactic knowledge, semantic knowledge, and syntactic/semantic associations \citep[Sections 9 and 10]{devt_sp_machine}.

If or when these problems are solved, there is potential for programming the SP system using written or spoken natural language, in much the same way that people can be given written or spoken instructions. However, achieving human levels of understanding is an ambitious goal and is not likely to be realised soon.

\section{Bringing `design' closer to `implementation'}\label{design_implementation_closer_section}

It has been established for some time that, in conventional development of software, one should begin with a relatively abstract high-level design (which is often represented graphically) and then translate that into a working program. There seem to be three main reasons for this approach:

\begin{itemize}

    \item With any kind of design, it is often useful to establish a relatively abstract ``big picture'' before filling in details.

    \item For the kinds of reasons described in Section \ref{non-automatic_programming_software_section}, it may be useful to disguise the details of a program behind syntactic sugar that is more congenial for programmers.

    \item Even with `high' level programming languages such as C++, Python, or Java, or `declarative' systems such as Prolog, it is often necessary to pay attention to the details of how the underlying machine will run a program, details that are not relevant to the more abstract `design' of the software, with its focus on entities and processes that are significant for the user.

\end{itemize}

The SP system probably makes no difference to the first and second of the above points, but it is likely to be helpful with the third. This is because, in the manner of declarative programming systems, it will probably allow programmers to specify `what' computations are to be achieved, and to reduce or eliminate the need to consider `how' the computations should be done.

\section{Possible reductions in the need for operations like compiling or interpretation}\label{no_compiling_or_interpretation_section}

At first sight, the SP system eliminates the need for anything like compiling or interpretation. This is because it works entirely via searches for full or partial matches between SP-patterns, or parts of SP-patterns, with corresponding unifications.

But it is likely that, in the development of the SP Machine, indexing will be introduced to record the first match between a given SP-symbol and any other SP-symbol, and thus speed up the later retrieval of the zero or more matching partners of the given SP-symbol \citep[Section 3.4]{devt_sp_machine}. And it is likely that similar measures will be introduced into the computer model for SP-neural \citep[Section 13.2]{devt_sp_machine}, a version of the SP Theory expressed in terms of neurons and their interconnections.

Indexing of that kind is similar in some respects to the use of compiling or interpretation in a conventional computing system. Hence it would be misleading to suggest that the SP system would eliminate the need for such operations. But there are potential gains in this area, especially if, at some later stage, it became feasible to introduce very fast and highly-parallel searching for matches between SP-patterns which may reduce or eliminate the need for indexing.

\section{Verification}\label{verification_section}

The SP system has potential to reduce the need for `verification' of software---meaning the process of checking that a software system meets its specifications---and there is corresponding potential for improvements in the quality of software. The main reasons for these potential benefits are:

\begin{itemize}

    \item {\em The potential of the system for automatic or semi-automatic creation of software} (Sections \ref{full_partial_automation_se_section} and \ref{semi-full_partial_automation_se_section}). To the extent that automatic or semi-automatic creation of software is possible, it should reduce or eliminate human-induced errors in software.

    \item {\em Potential reductions in the sizes of software systems}. The potential of the system for reductions in the overall sizes of software systems (Section \ref{overall_simplification_of_applications_section}) means that there are likely to be fewer opportunities to introduce bugs into software, and, probably, less searching would be required in the detection of bugs via static analysis of software.

    \item {\em Bringing `design' closer to `implementation'}. To the extent that `design' and `implementation' may be merged (Section \ref{design_implementation_closer_section}), and in particular to the extent that SP software may concentrate on `what' the user needs and reduce or eliminate details of `how' the underlying machine may meet those needs, there is potential to reduce the numbers of bugs in programs.

\end{itemize}

\section{Validation}\label{validation_section}

In addition to its potential with verification, the SP system has potential to strengthen the process of ``validation'' in software development---meaning the process of checking that a software system fulfills its intended purpose.

As with verification, the potential of the SP system for the automatic or semi-automatic creation of software means elimination or reduction of the kinds of human error that may send a program off track.

Also, the potential of the SP system to bring `design' and `implementation' closer together (Section \ref{design_implementation_closer_section}) can mean fewer opportunities for a program to drift away from its original conception.

\section{Seamless integration of `software' with `database'}\label{integration_of_software_with_database_section}

In the SP system, {\em all} kinds of knowledge are represented with arrays of atomic {\em SP-symbols} in one or two dimensions (Appendix \ref{overview_appendix}), and {\em all} kinds of processing is achieved via the matching and unification of SP-patterns. For these two reasons, and because of the system's potential for {\em universal artificial intelligence} (UAI) (Appendix \ref{uai_appendix}), there would be no distinction in the SP system between `software' and `database', as there is normally in conventional software engineering projects.

A potential benefit of this kind seamless integration of software and database is elimination of awkward incompatibilities between different kinds of knowledge and elimination of the need for translations where incompatibilities exist.

\section{An overall simplification of computing applications}\label{overall_simplification_of_applications_section}

With the SP system, there is potential for an overall simplification of applications compared with what is required with ordinary computers \citep[Section 5]{sp_benefits_apps}.

\begin{figure}[!htbp]
\centering
\includegraphics[width=0.9\textwidth]{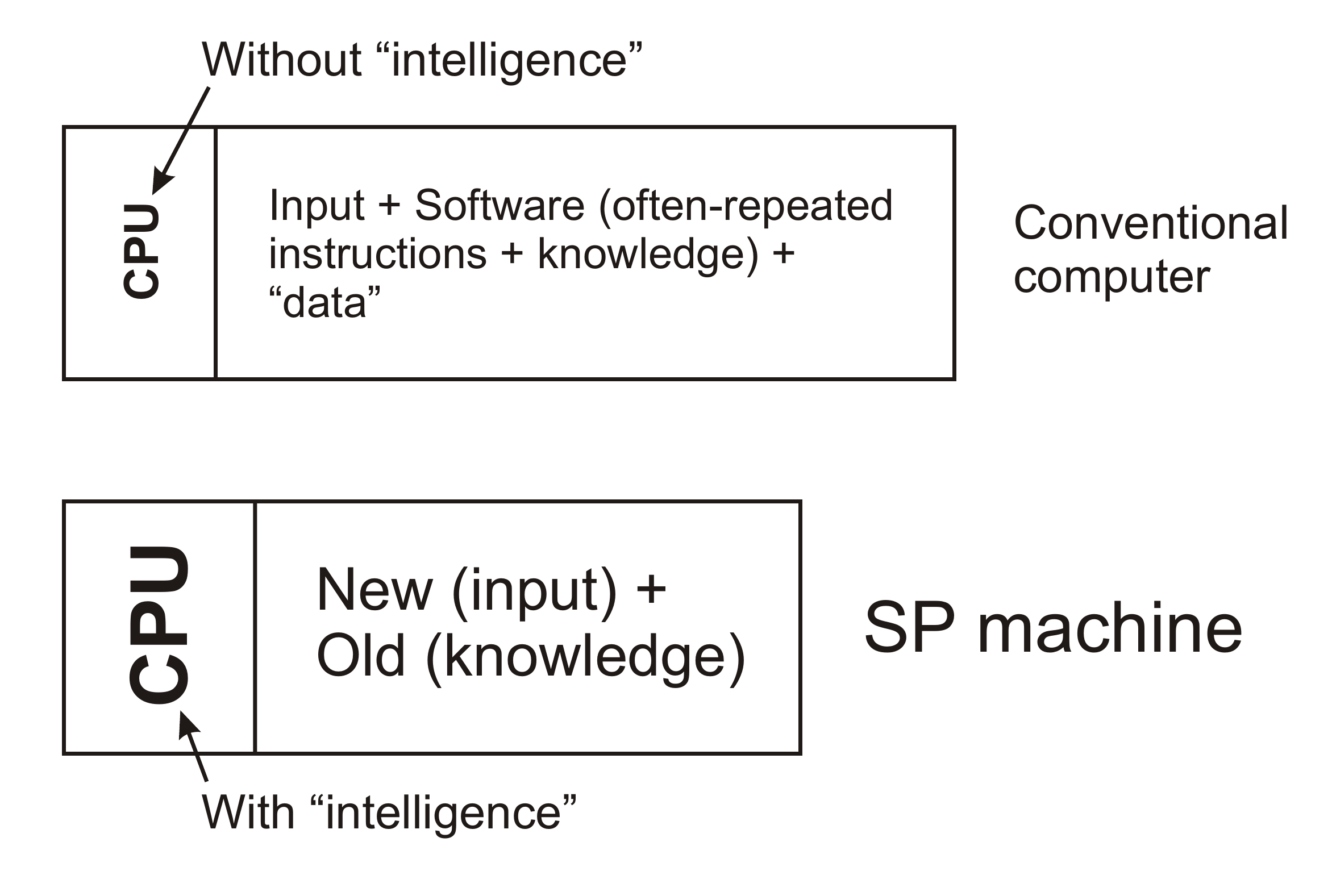}
\caption{Schematic representations of a conventional computer and the proposed {\em SP machine}, showing potential benefits in terms of simplification, as discussed in the text. Adapted from Figure 4.7 in \protect\cite{wolff_2006}, with permission.}
\label{two_schematic_computers_figure}
\end{figure}

In broad terms, this potential arises because of the way in which conventional software contains often-repeated procedures for searching amongst data, and similar `low level' operations needed to overcome shortcomings in conventional CPUs. In an SP system, the `CPU' is relatively complex but with fewer of the shortcomings of conventional CPUs, so that that relative complexity is, probably, more than offset by simplifications in software with data, as shown schematically in Figure \ref{two_schematic_computers_figure}. That relative advantage is likely to grow, roughly in proportion to the numbers of applications and their sizes.

This kind of idea is not new. In the early days of databases, each database had its own procedures for searching and for retrieval of information, and it had its own user interface and procedures for printing, and so on. People soon realised that it would make better sense to develop a general-purpose system for the management of data, with a user interface and system for retrieval of data, and to load it with different bodies of data according to need. There was a similar evolution in expert systems, from bespoke systems to general-purpose `shells'.

\section{Reducing the variety of formats and formalisms in computing}\label{reducing_formats_and_formalisms_section}

As noted in Section \ref{integration_of_software_with_database_section}, the SP system has potential for {\em universal artificial intelligence} (UAI). What this means in the SP programme of research, and how the concept of a UAI differs from alternatives such as the concept of a universal Turing machine, is discussed in Appendix \ref{uai_appendix}.

If indeed this expectation is born out, and the evidence is strong, there is clear potential for use of the SP Machine to clean up the curse of variety in the thousands of different formats and formalisms which exist for the representation of data, and the hundreds of different computer languages for describing how data may be processed (Appendix \ref{the_curse_of_variety_appendix}).

\section{Version control}\label{version_control_section}

In a typical software engineering project, there is a need to keep track of the parts and sub-parts of the developing program. At the same time, there is a need to keep track of a hierarchy of versions and subversions. And, associated with each part or version, there may be several different kinds of document, including a statement of requirements, a high-level design, a low-level design, and notes. To avoid awkward inconsistencies, these things should be smoothly integrated.\footnote{The problem of integrating a class-inclusion hierarchy with a part-whole hierarchy---a problem that arose in connection with the development of an ``Integrated Project Support Environment'' (IPSE) when I was working as a software engineer with Praxis Systems plc---was one of the main sources of inspiration for the development of the SP system.}

The SP system provides a neat solution to the problem of integrating a class-inclusion hierarchy with a part-whole hierarchy, as described in \citet[Section 9.1]{sp_extended_overview} and \citet[Section 6.4]{wolff_2006}.\footnote{The solution also applies to class-inclusion heterarchies, meaning a class-inclusion hierarchy with cross-classification.} Although these sources do not demonstrate the point, it appears that the SP system also provides for each version or part to have one or more associated documents, as outlined above.

Also relevant to these issues is a brief discussion of how to maintain multiple versions and parts of a document or web page in \citet[Section 6.10.3]{sp_benefits_apps}.

\section{Technical debt}\label{technical_debt_section}

As noted in \citet[Section 6.6.6]{sp_benefits_apps}, the SP system has potential to reduce or eliminate the problem of `technical debt', meaning the way in which software systems can become progressively more unmanageable with the passage of time, owing to an accumulation of postponed or abandoned maintenance tasks, or a progressive deterioration in the design quality or maintainability of the software via the repeated application of `fixes' in response to short-terms concerns, without sufficient attention to their global and long-term effects.

The SP system may reduce or eliminate the problem of technical debt by streamlining the process of software development via automatic or semi-automatic automation of software development, by reducing the gap between design and implementation, by streamlining processes of verification and validation, and other facilitations described in preceding sections.

\section{Conclusion}

This paper describes a novel approach to software engineering derived from the {\em SP Theory of Intelligence} and its realisation in the {\em SP Computer Model}. It is anticipated that the SP Theory and the SP Computer Model, together, will be the basis for the development of an industrial-strength {\em SP Machine}. And a mature version of the SP Machine is seen as the likely vehicle for software engineering as described in this paper.

Although concepts associated with software engineering may seem far removed from the structure and workings of the SP system, many of those concepts map quite neatly into elements of the SP system (Section \ref{relating_conventional_concepts_to_sp_concepts}).

Potential benefits of this new approach to software engineering include:

\begin{itemize}

    \item {\em The automation of semi-automation of software development}. Taking advantage of the SP system's strengths and potential in unsupervised learning, there is clear potential for the automation or semi-automation of software development.

    \item {\em Non-automatic programming of the SP system}. Where it is not possible to create software automatically, or when human assistance is needed, there is clear potential for programming the SP system in much the same way as a conventional system.

    \item {\em Programming via natural language}. An ambitious goal, which is not likely to be realised soon, is to bring the SP system to a point where it has human levels of understanding and production of natural language, so that the SP system may be `programmed' in much the same way that people can be given written or spoken instructions.

    \item {\em Reducing or eliminating the distinction between `design' and `implementation'}. By contrast with conventional systems, there is potential in the SP system to reduce or eliminate the distinction between `design' and `implementation'. This is because aspects of software design, such as structured programming and object-oriented programming, may be expressed directly with SP-patterns.

    \item {\em Reducing or eliminating operations like compiling or interpretation}. The SP system has potential to reduce or eliminate operations like compiling or interpretation. This is because the system works directly on `source' code by searching for patterns or parts of patterns that match each other. But it seems likely that indexing of matches between SP-symbols will speed up the system, and the compiling of such an index may be seen to be similar to what is entailed in conventional compiling or interpretation.

    \item {\em Reducing or eliminating the need for verification of software}. The need for verification of SP software may be reduced or eliminated: via the potential of the SP system for the automatic or semi-automatic creation of software; because compression of software is likely to reduce the opportunities for bugs to be introduced; and because there is likely to be a reduced need to bridge the divide between design and implementation.

    \item {\em Reducing or eliminating the need for validation of software}. The SP system also has potential to help ensure that software fulfills its intended purpose. This is because of the system's potential for automatic or semi-automatic creation of software and because of the way in which design and implementation may be brought closer together or merged.

    \item {\em No formal distinction between program and database}. Unlike conventional systems, where `programs' and `databases' are distinguished quite sharply, there is no formal distinction of that kind in the SP system because all kinds of knowledge are expressed with SP-patterns. This can mean useful simplifications on occasion, and it can reduce or remove awkward incompatibilities.

    \item {\em Potential for an overall simplification of computing applications}. Despite the fact that the processing `core' of the SP system is, almost certainly, more complex than the CPU of a conventional computer, there is potential with the SP system for an overall simplification of computing applications, when hardware and software are considered together.

    \item {\em Potential for substantial reductions in the number of types of data file and the number of computer languages}. Because of the SP system's potential as a {\em universal artificial intelligence} (UAI), there is potential to reduce the many thousands of types of data file to one, and to reduce the hundreds of different computer languages to one.

    \item {\em Allowing programmers to concentrate on `world-oriented' parallelism, without worries about parallelism to speed up processing}. With a mature version of the SP Machine, it is intended that parallelism that is designed only for the purpose of speeding up processing will be built into the system, so that programmers need not worry about it. They would be free to concentrate on parallelism in the real world, perhaps with assistance from unsupervised learning.

    \item {\em Benefits for version control}. The SP system has potential to help organise all the knowledge associated with any given software development project, with provision for: the representation of versions  of the software; the representation of parts and sub-parts of the software; the seamless integration of version hierarchies with part-whole hierarchies; and, for any given version or part, the representation of the one or more kinds of information associated with that element. It provides for cross-classification where that is required.

    \item {\em Reducing technical debt}. The potential of the SP system to increase the efficiency of software development can mean reductions in `technical debt', meaning the way in which software systems can become progressively more unmanageable with the passage of time, owing to short-term fixes and the postponement or abandonment of maintenance tasks.

\end{itemize}


\newpage

\section*{Appendices}

\appendix

\section{Outline of the SP system}\label{outline_of_sp_system_appendix}

To help ensure that this paper is free standing, the SP system is described here in outline with enough detail to make the rest of the paper intelligible.

The {\em SP Theory of Intelligence} and its realisation in the {\em SP Computer Model} is the product of a unique extended programme of research aiming to simplify and integrate observations and concepts across artificial intelligence, mainstream computing, mathematics, and human learning, perception, and cognition, with information compression as a unifying theme.\footnote{This ambitious objective is in keeping with Occam's Razor. And as a means of solving the exceptionally difficult problem of developing general, human-level artificial intelligence, it is in keeping with ``If a problem cannot be solved, enlarge it'', attributed to President Eisenhower; it chimes with Allen Newell's exhortation that psychologists should work to understand ``a genuine slab of human behaviour'' \citep[p.~303]{newell_1973} and his work on {\em Unified Theories of Cognition} \citep{newell_1990}; and it is in keeping with the quest for ``Artificial General Intelligence'' ({\em Wikipedia}, \href{http://bit.ly/1ZxCQPo}{bit.ly/1ZxCQPo}, retrieved 2017-08-15).}

\sloppy The latest version of the SP Computer Model is SP71. Details of where the source code and associated files may be obtained are here: \href{http://www.cognitionresearch.org/sp.htm\#ARCHIVING}{www.cognitionresearch.org/sp.htm\#ARCHIVING}.

It is envisaged that the SP Computer Model will provide the basis for the development of an industrial-strength {\em SP Machine}, described briefly in Appendix \ref{sp_machine_appendix}, below.

\sloppy The SP system is described most fully in \citet{wolff_2006} and quite fully but more briefly in \citet{sp_extended_overview}. Other publications from this programme of research are detailed, many with download links, on \href{http://www.cognitionresearch.org/sp.htm}{www.cognitionresearch.org/sp.htm}.

\subsection{Overview}\label{overview_appendix}

\begin{figure}[!htbp]
\centering
\includegraphics[width=0.6\textwidth]{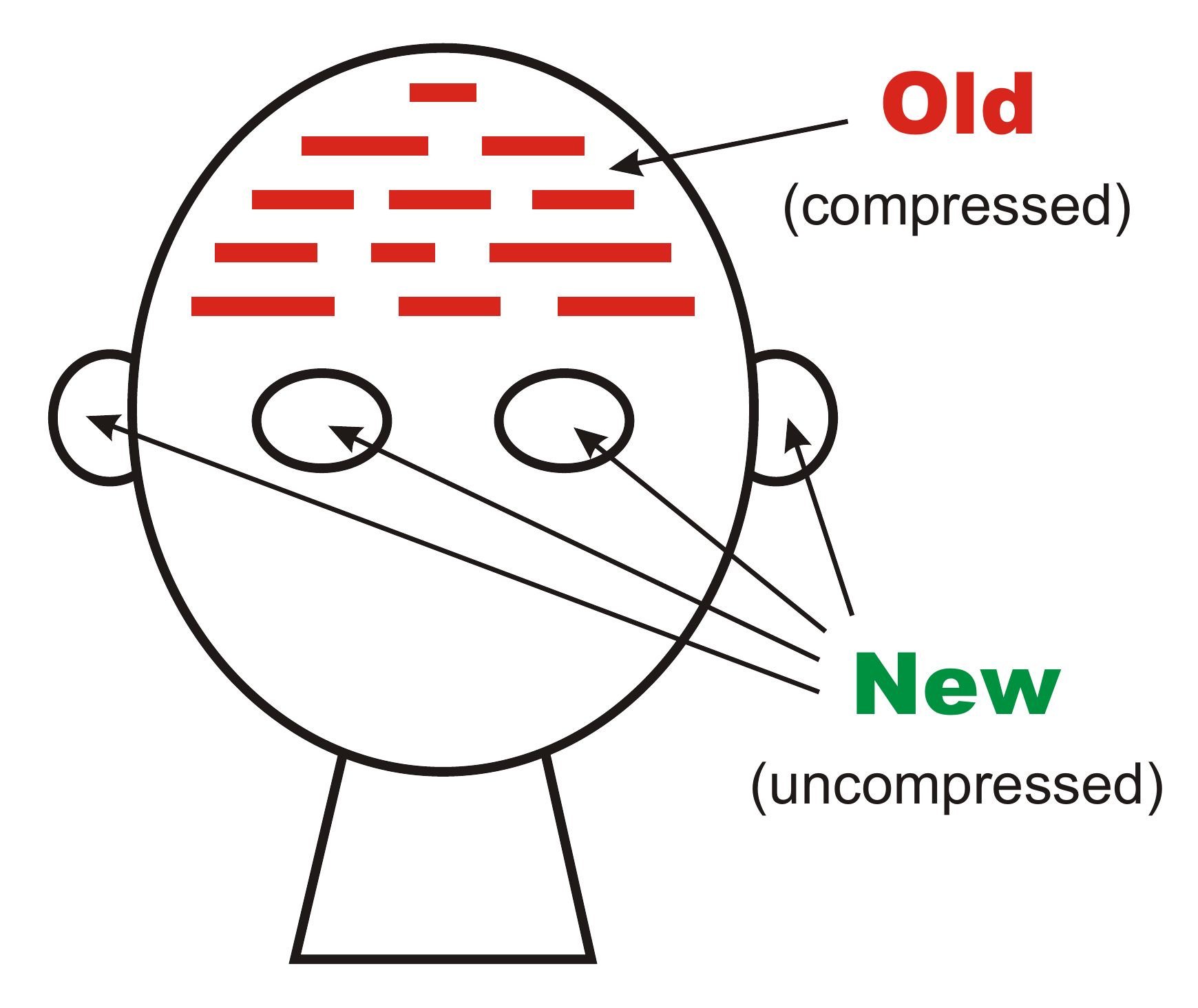}
\caption{Schematic representation of the SP system from an `input' perspective. Reproduced with permission from Figure 1 in \protect\citet{sp_extended_overview}.}
\label{sp_input_perspective_figure}
\end{figure}

The SP Theory is conceived as a brain-like system which receives {\em New} information via its senses and stores some or all of it in compressed form as {\em Old} information, as shown schematically in Figure \ref{sp_input_perspective_figure}.

Both New and Old information are expressed as arrays of atomic {\em SP-symbols} in one or two dimensions called {\em SP-patterns}. To date, the SP Computer Model works only with one-dimensional SP-patterns but it is envisaged that it will be generalised to work with two-dimensional SP-patterns.

In this context, an `SP-symbol' is simply a mark that can be matched with any other SP-symbol to determine whether they are the same or different. No other result is permitted. Apart from some distinctions needed for the internal workings of the SP system, SP-symbols do not have meanings such as `plus' ('$+$'), `multiply' (`$*$'), and so on. Any meaning associated with an SP-symbol derives entirely from other SP-symbols with which it is associated.

\subsection{Multiple sequemce alignments in bioinformatics}\label{multiple_alignments_in_bioinformatics_appendix}

At the heart of the SP system is information compression via the matching and unification of patterns (ICMUP). More specifically, a central part of the SP system is a concept of {\em SP-multiple-alignment}, borrowed and adapted from the concept of `multiple sequence alignment' in bioinformatics.

The original concept is an arrangement of two or more DNA sequences or sequences of amino acid residues, in rows or columns, with judicious `stretching' of selected sequences in a computer to bring symbols that match each other from row to row, as many as possible, into line. An example of such a multiple sequennce alignment of five DNA sequences is shown in Figure \ref{DNA_figure}.

\begin{figure}[!htbp]
\fontsize{09.00pt}{10.80pt}
\centering
{\bf
\begin{BVerbatim}
  G G A     G     C A G G G A G G A     T G     G   G G A
  | | |     |     | | | | | | | | |     | |     |   | | |
  G G | G   G C C C A G G G A G G A     | G G C G   G G A
  | | |     | | | | | | | | | | | |     | |     |   | | |
A | G A C T G C C C A G G G | G G | G C T G     G A | G A
  | | |           | | | | | | | | |   |   |     |   | | |
  G G A A         | A G G G A G G A   | A G     G   G G A
  | |   |         | | | | | | | |     |   |     |   | | |
  G G C A         C A G G G A G G     C   G     G   G G A
\end{BVerbatim}
}
\caption{A `good' multiple sequence alignment amongst five DNA sequences. Reproduced with permission from Figure 3.1 in \citet{wolff_2006}.}
\label{DNA_figure}
\end{figure}

\subsection{SP-multiple-alignments in the SP system}\label{sp_multiple_alignment_appendix}

In the SP system, multiple alignments are sufficiently different from those in bioinformatics for them to be given a different name: {\em SP-multiple-alignments}.\footnote{This name has been introduced fairly recently to make clear that there are important differences between the two kinds of multiple alignment.} The distinctive features of an SP-multiple-alignment are:

\begin{itemize}

    \item One New SP-pattern is shown in row 0 (or column 0 when SP-patterns are arranged in columns).\footnote{Sometimes there is more than one New SP-pattern in row 0 or column 0.}

    \item The Old SP-patterns are shown in the remaining rows (or columns), one SP-pattern per row (or column).

    \item As with the original concept of multiple alignment, the aim in building multiple alignments is to bring matching SP-symbols into alignment. More specifically in SP-multiple-alignments, the aim is to create or discover one or more `good' SP-multiple-alignments that allow the New SP-pattern to be encoded economically in terms of the Old SP-patterns. How this encoding is done is described in \citet[Section 2.5]{wolff_2006} and in \citet[Section 4.1]{sp_extended_overview}.

\end{itemize}

\begin{figure}[!htbp]
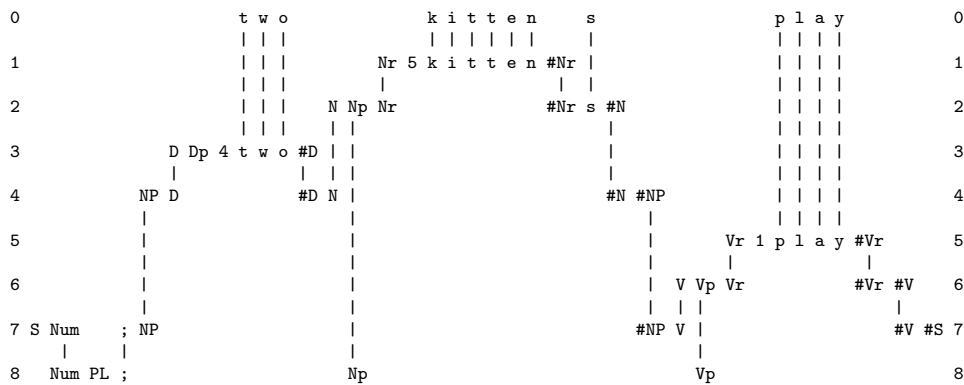

\fontsize{07.00pt}{08.40pt}
\centering
{\bf
\begin{BVerbatim}
0                      t w o              k i t t e n     s                  p l a y           0
                       | | |              | | | | | |     |                  | | | |
1                      | | |         Nr 5 k i t t e n #Nr |                  | | | |           1
                       | | |         |                 |  |                  | | | |
2                      | | |    N Np Nr               #Nr s #N               | | | |           2
                       | | |    | |                         |                | | | |
3               D Dp 4 t w o #D | |                         |                | | | |           3
                |            |  | |                         |                | | | |
4            NP D            #D N |                         #N #NP           | | | |           4
             |                    |                             |            | | | |
5            |                    |                             |       Vr 1 p l a y #Vr       5
             |                    |                             |       |             |
6            |                    |                             |  V Vp Vr           #Vr #V    6
             |                    |                             |  | |                   |
7 S Num    ; NP                   |                            #NP V |                   #V #S 7
     |     |                      |                                  |
8   Num PL ;                      Np                                 Vp                        8
\end{BVerbatim}
}
\caption{The best SP-multiple-alignment created by the SP Computer Model with a store of Old SP-patterns like those in rows 1 to 8 (representing grammatical structures, including words) and a New SP-pattern (representing a sentence to be parsed) shown in row 0. Adapted with permission from Figures 1 in \protect\citet{wolff_sp_intelligent_database}.}
\label{parsing_figure}
\end{figure}

An example of an SP-multiple-alignment is shown in Figure \ref{parsing_figure}.

In this SP-multiple-alignment, a sentence is shown as a New SP-pattern in row 0. The remaining rows show Old SP-patterns, one per row, representing grammatical structures including words. The overall effect is to analyse (parse) the sentence into its parts and subparts. The SP-pattern in row 8 shows the association between the plural subject of the sentence, marked with the SP-symbol `\texttt{Np}', and the plural main verb, marked with the SP-symbol `\texttt{Vp}'.

Because, with most ordinary multiple sequence alignments or with SP-multiple-alignments, there is an astronomically large number of ways in which sequences may be aligned, discovering good multiple alignments means the use of heuristic methods: building each multiple alignment in stages and discarding all but the best few multiple alignment at the end of each stage. With this kind of technique it is normally possible to find multiple alignments that are reasonably good but it is not normally possible to guarantee that the best possible multiple alignment has been found.

The concept of SP-multiple-alignment has proved to be extraordinarily powerful: in aspects of intelligence (Appendix \ref{versatility_in_intelligence_appendix}), in the representation of knowledge (Appendix \ref{versatility_in_rk_appendix}), and in the seamless integration of diverse aspects of intelligence and diverse kinds of knowledge in any combination (Appendix \ref{seamless_integration_appendix}). It could prove to be as significant for an understanding of intelligence as is DNA for biological sciences: it could be the `double helix' of intelligence.

\subsection{Unsupervised learning}\label{unsupervised_learning_appendix}

Unsupervised learning in the SP system is described quite fully in \citet[Sections 3.9 and 9.2]{wolff_2006}. The aim with unsupervised learning in the SP system is, for a given set of New SP-patterns, to create one or two {\em grammars}---meaning collections of Old SP-patterns---that are effective at encoding the given set of New SP-patterns in an economical manner.

The process of creating good grammars entails the creation of Old SP-patterns, partly by the direct assimilation of New SP-patterns and partly via the building of SP-multiple-alignments---which provides a means of creating Old SP-patterns and via the splitting of New SP-patterns and the splitting of pre-existing Old SP-patterns.

The building of SP-multiple-alignments also provides a means of evaluating candidate grammars in terms of their effectiveness at encoding the given set of New SP-patterns in an economical manner.

As with the building of SP-multiple-alignments, the creation of good grammars requires heuristic search through the space of alternative grammars: creating grammars in stages and discarding low-scoring grammars at the end of each stage.

The SP Computer Model can discover plausible grammars from samples of English-like artificial languages. This includes the discovery of segmental structures, classes of structure, and abstract SP-patterns.

At present, the program has two main weaknesses outlined in \citet[Section 3.3]{sp_extended_overview}: it does not learn intermediate levels of abstraction or discontinuous dependencies in data. However, it appears that these problems are soluble, and it seems likely that their solution would greatly enhance the performance of the system in the learning of diverse kinds of knowledge.

To ensure that unsupervised learning in the SP system is robust and useful across a wide range of different kinds of data, it will be necessary for the system, including its procedures for unsupervised learning, to have been generalised for two-dimensional SP-patterns as well as one-dimensional SP-patterns (Appendix \ref{overview_appendix}).

\subsection{The SP Machine}\label{sp_machine_appendix}

As mentioned earlier, it is envisaged that an industrial-strength {\em SP Machine} will be developed from the SP Theory and the SP Computer Model \citep{devt_sp_machine}. Initially, this will be created as a high-parallel software virtual machine, hosted on an existing high-performance computer. An interesting possibility is to develop the SP Machine as a software virtual machine that is driven by the high-parallel search processes in any of the leading internet search engines.

Later, there may be a case for developing new hardware for the SP Machine, to take advantage of optimisations that may be achieved by tailoring the hardware to the characteristics of the SP system. In particular, there is potential for substantial gains in efficiency and savings in energy compared with conventional computers by taking advantage of statistical information that is gathered by the SP system as a by-product of how it works (\citet[Section IX]{sp_big_data}, \citet[Section III]{sp_autonomous_robots}, \citet[Section 14]{devt_sp_machine}).

\begin{figure}[!htbp]
\centering
\includegraphics[width=0.9\textwidth]{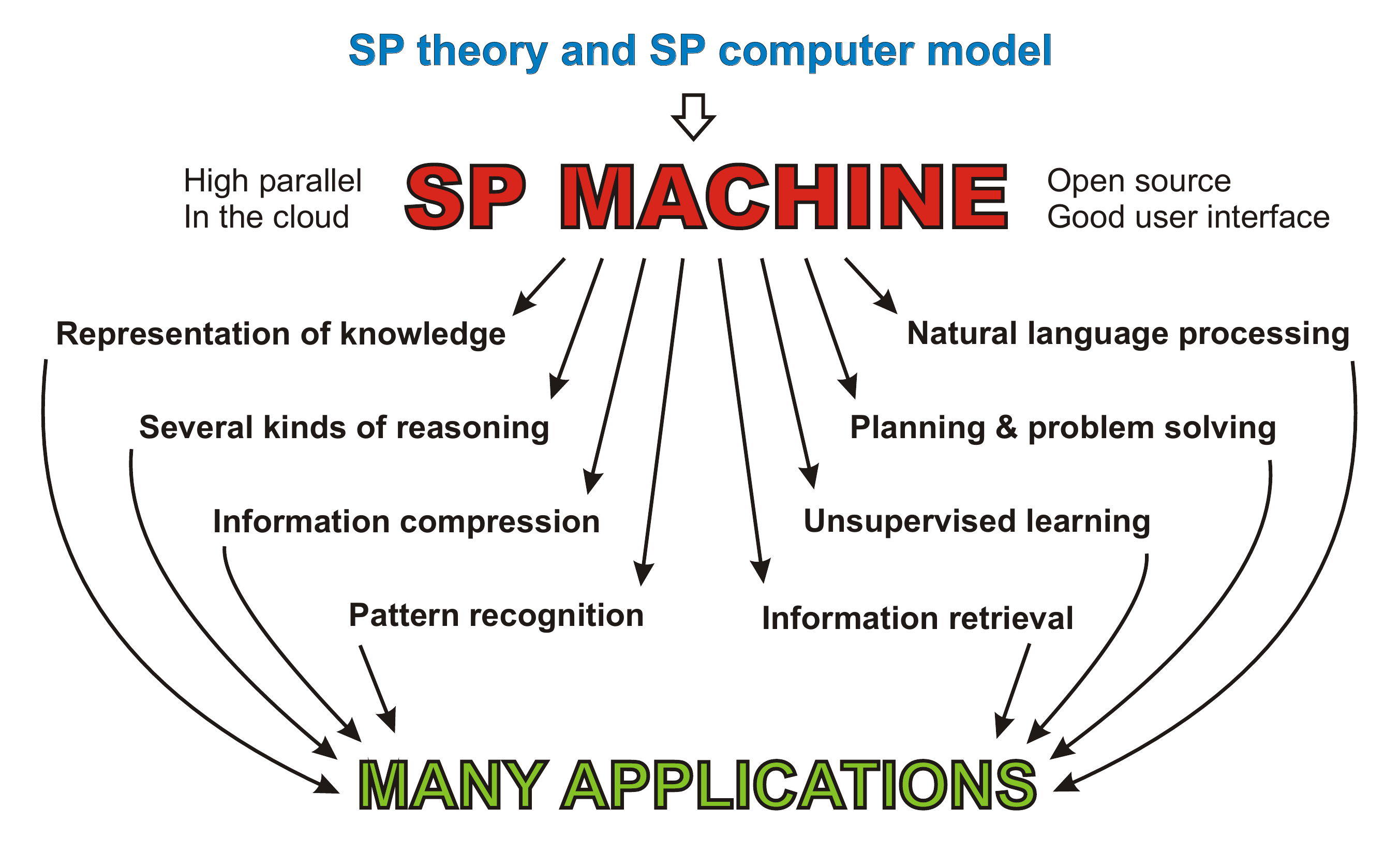}
\caption{Schematic representation of the development and application of the SP machine. Reproduced from Figure 2 in \cite{sp_extended_overview}, with permission.}
\label{sp_machine_figure}
\end{figure}

A schematic representation of how the SP Machine may be developed and applied is shown in Figure \ref{sp_machine_figure}.

\subsection{Distinctive features and advantages of the SP system}\label{distinctive_features_advantages_appendix}

Distinctive features of the SP system and its main advantages compared with AI-related alternatives are described in \citet{sp_alternatives}. In particular, Section V of that paper describes thirteen problems with deep learning in artificial neural networks and how, with the SP system, those problems may be overcome. The SP system also provides a comprehensive solution to a fourteenth problem with deep learning---``catastrophic forgetting''---meaning the way in which new learning in a deep learning system wipes out old memories.\footnote{A solution has been proposed in \citet{kirkpatrick_2017} but it appears to be partial, and it is unlikely to be satisfactory in the long run.}

The main strengths of the SP system are in its versatility in the representation of several kinds of knowledge (Appendix \ref{versatility_in_rk_appendix}), its versatility in several aspects of intelligence (Appendix \ref{versatility_in_intelligence_appendix}), and because these things all flow from one relatively simple framework---the SP-multiple-alignment concept---they may work together seamlessly in any combination (Appendix \ref{seamless_integration_appendix}). That kind of seamless integration appears to be essential in any system that aspires to general human-level artificial intelligence.

\subsection{Potential benefits and applications of the SP system}

Potential benefits and applications of the SP system are described in several peer-reviewed papers, copies of which may be obtained via links from \href{http://www.cognitionresearch.org/sp.htm}{www.cognitionresearch.org/sp.htm}: the SP system may help to solve nine problems with big data \citep{sp_big_data}; it may help in the development of human-like intelligence in autonomous robots \citep{sp_autonomous_robots}; the SP system may help in the understanding of human vision and in the development of computer vision \citep{sp_vision}; it may function as a database system with intelligence \citep{wolff_sp_intelligent_database}; it may assist medical practitioners in medical diagnosis \citep{wolff_medical_diagnosis}; it provides insights into commonsense reasoning \citep{sp_csrk}; and it has several other potential benefits and applications described in \citet{sp_benefits_apps}. And, of course, this paper describes how the SP system may be applied in software engineering.

\section{Towards universal artificial intelligence (UAI)}\label{uai_appendix}

In Sections \ref{integration_of_software_with_database_section} and \ref{reducing_formats_and_formalisms_section}, it has been noted that the SP system has potential for {\em universal artificial intelligence} (UAI). The purpose of this appendix is to describe what this means and to distinguish the concept from alternatives such as a `universal Turing machine' (UTM) \citep{turing_1936}.

The idea that something may have UAI or be a UAI derives from the concept of a {\em universal framework for the representation and processing of diverse kinds of knowledge} (UFK) \citep[Section III]{sp_big_data} but gives weight to the concept of (human-like) `intelligence'.

The idea that the SP system has potential for UAI may at first sight seem to be redundant since it has been recognised for some time that all kinds of computing may be understood in terms of the workings of a UTM or ideas which are recognised as equivalent such as Post's `canonical system' \citep{post_1943}, or Church's `lambda calculus' \citep{church_1941}, or indeed the many conventional computers that are in use today. For the sake of brevity these will be referred to collectively as CCs, short for `conventional computers'.

The suggestion here is that, by definition: 1) a UAI should demonstrate human-like intelligence, 2) it should be able to represent any kind of knowledge, 3) it should provide for any kind of processing within the limits set by computational complexity, 4) it should facilitate the seamless integration of diverse kinds of knowledge and diverse kinds of processing in any combination, and 5) it should do these things efficiently. In short, a UAI is a Turing-equivalent device with human-like intelligence.

The potential of the SP system in areas 1), 2), 4), and 5), and how it differs from a CC, is described in the following four subsections.

\subsection{Versatility in aspects of intelligence via the powerful concept of SP-multiple-alignment}\label{versatility_in_intelligence_appendix}

As noted in Appendix \ref{sp_multiple_alignment_appendix}, the concept of SP-multiple-alignment has the potential to be the `double helix' of intelligence, the key to the versatility of the SP system in aspects of intelligence, summarised here:

\begin{itemize}

    \item {\em Unsupervised learning via the processing of SP-multiple-alignments}. The SP system has strengths and potential in `unsupervised' learning of new knowledge, meaning learning without the assistance of a `teacher' or anything equivalent. As outlined in Appendix \ref{unsupervised_learning_appendix}, unsupervised learning is achieved in the SP system via the processing of SP-multiple-alignments to create Old SP-patterns, directly and indirectly, from New SP-patterns, and to build collections of Old SP-patterns, called {\em grammars} which are relatively effective in the compression of New SP-patterns (\citet[Chapter 9]{wolff_2006}, \citet[Section 5]{sp_extended_overview}).

        Unsupervised learning appears to be the most fundamental form of learning, with potential as a foundation for other forms of learning such as reinforcement learning, supervised learning, learning by imitation, and learning by being told.

    \item {\em How other aspects of intelligence flow from the building of SP-multiple-alignments}. By contrast with the way in which the SP system models unsupervised learning via the processing of already-constructed `good' SP-multiple-alignments, other aspects of intelligence derive from the building of SP-multiple-alignments (Appendix \ref{sp_multiple_alignment_appendix}). These other aspects of intelligence include: analysis and production of natural language; pattern recognition that is robust in the face of errors in data; pattern recognition at multiple levels of abstraction; computer vision \citep{sp_vision}; best-match and semantic kinds of information retrieval; several kinds of reasoning (more under the next bullet point); planning; and problem solving (\citet[Chapters 5 to 8]{wolff_2006}, \citet[Sections 5 to 14]{sp_extended_overview}).

    \item {\em How several kinds of reasoning flow from the building of SP-multiple-alignments}. In scientific research and in the applications of science, what is potentially one of the most useful attributes of the SP system is its versatility in reasoning, described in \citet[Chapter 7]{wolff_2006} and \citet[Section 10]{sp_extended_overview}. Strengths of the SP system in reasoning, derived from the building of SP-multiple-alignments, include: one-step `deductive' reasoning; chains of reasoning; abductive reasoning; reasoning with probabilistic networks and trees; reasoning with `rules'; nonmonotonic reasoning and reasoning with default values; Bayesian reasoning with `explaining away'; causal reasoning; reasoning that is not supported by evidence; the already-mentioned inheritance of attributes in class hierarchies; and inheritance of contexts in part-whole hierarchies. There is also potential for spatial reasoning \citep[Section IV-F.1]{sp_autonomous_robots}, and for what-if reasoning \citep[Section IV-F.2]{sp_autonomous_robots}.

\end{itemize}

\subsubsection{Generality in artificial intelligence}\label{generality_in_ai_appendix}

The close connection that is known to exist between information compression and concepts of prediction and probability \citep{solomonoff_1964,solomonoff_1997,li_vitanyi_2014}, the central role of information compression in the SP-multiple-alignment framework (Appendix \ref{sp_multiple_alignment_appendix}), and the versatility of the SP-multiple-alignment framework in the representation of knowledge (Appendix \ref{versatility_in_rk_appendix}) and aspects of intelligence (Appendix \ref{versatility_in_intelligence_appendix}), suggest that SP-multiple-alignment may prove to be the key to the development of general, human-like artificial intelligence.

\subsubsection{What about things that the SP system can't do, except with some kind of `programming' or `training'?}\label{sp_programming_training_appendix}

In considering the possibility that the SP system might be developed into a UAI is that, while the mechanisms for the building and processing of SP-multiple-alignments, yield several different AI-related capabilities, described above, there are lots of things that a newly-created system, without any `experience', would not be able to do. It would not, for example, have any knowledge of how to hold a pencil, how to climb a ladder, how to negotiate an international treaty, and so on.

Is it reasonable to suggest that such a system might be a UAI when there there are so many shortcomings in what it can do? The answer, of course, is ``Yes, such a system can be `universal' in exactly the same way that a universal Turing machine, or a newborn baby, is universal'', because in all three cases there is the potential to do a wide variety of different things, provided that it has appropriate knowledge, acquired via learning (babies and AI systems) or programming (computers).

Since procedures or processes are forms of knowledge, and since we have reason to believe that the SP system may accommodate any kind of knowledge (Appendix \ref{versatility_in_rk_appendix}), it is reasonable to believe that the SP system may in principle, with the right knowledge, do any kind of computation that is not ruled out by over-large computational complexity.

\subsubsection{How the SP system differs from a CC in aspects of intelligence}\label{sp_cc_differences_in_intelligence_appendix}

With regard to the modelling of human-like intelligence, the main attraction of the SP system compared with CCs, is its versatility in diverse aspects of intelligence (Appendix \ref{versatility_in_intelligence_appendix}) and its potential for the seamless integration of diverse aspects of intelligence and diverse kinds of knowledge, in any combination (Appendix \ref{seamless_integration_appendix}).

Unless a CC has been specifically programmed with SP capabilities---in which case it would be an SP system, not a CC---it would be lacking in the above-mentioned capabilities, and, arguably, for that reason, is likely to fall short of general human-like artificial intelligence.

\subsection{Versatility in the representation of knowledge via the powerful concept of SP-multiple-alignment}\label{versatility_in_rk_appendix}

The SP system has the potential for UAI because, although SP-patterns are not very expressive in themselves, they come to life in the SP-multiple-alignment framework. Within that framework, they may serve in the representation of several different kinds of knowledge, including: the syntax of natural languages; class-inclusion hierarchies (with or without cross classification); part-whole hierarchies; discrimination networks and trees; if-then rules; entity-relationship structures \citep[Sections 3 and 4]{wolff_sp_intelligent_database}; relational tuples ({\em ibid}., Section 3), and concepts in mathematics, logic, and computing, such as `function', `variable', `value', `set', and `type definition' (\citet[Chapter 10]{wolff_2006}, \citet[Section 6.6.1]{sp_benefits_apps}).

With the addition of two-dimensional SP-patterns to the SP Computer Model, there is potential for the SP system to represent such things as: photographs; diagrams; structures in three dimensions \citep[Section 6.1 and 6,2]{sp_vision}; and procedures that work in parallel \citep[Sections V-G, V-H, and V-I, and Appendix C]{sp_autonomous_robots}.

\subsubsection{Generality in the representation of knowledge}\label{generality_in_rk_appendix}

The generality of information compression as a means of representing knowledge in a succinct manner, the central role of information compression in the SP-multiple-alignment framework, and the versatility of that framework in the representation of knowledge, suggest that SP-multiple-alignment may prove to be a means of representing {\em any} kind of knowledge, as would be needed if the SP system were to be a UAI.

\subsubsection{How the SP system differs from a CC in the representation of knowledge}\label{sp_cc_differences_in_rk_appendix}

With regard to the representation of knowledge, attractions of the SP system compared with CCs are:

\begin{itemize}

    \item The SP system provides for the succinct representation of knowledge via ICMUP and the powerful concept of SP-multiple-alignment. By contrast, information compression, ICMUP, and SP-multiple-alignments are barely recognised as guides or principles for the representation of knowledge in CCs.\footnote{For example, none of these ideas is mentioned in ``Knowledge representation and reasoning'', {\em Wikipedia}, \href{http://bit.ly/2fmKVtP}{bit.ly/2fmKVtP}, retrieved 2017-08-07.}

    \item The versatility of the SP system in the representation of knowledge is combined with some constraint---knowledge must be represented with SP-patterns and processed via the building and manipulation of SP-multiple-alignments (Appendix \ref{versatility_in_rk_appendix})---and that constraint seems to be largely responsible for how the system facilitates the seamless integration of different kinds of knowledge (Appendix \ref{seamless_integration_appendix}).

        By contrast, the representation of knowledge in a CC is a free-for-all: any kind of structure that may be represented with arrays $0$s and $1$s is accepted. This relative lack of discipline seems to be largely responsible for the excessive number of formats and formalisms in computing today (Appendix \ref{the_curse_of_variety_appendix}) and the many incompatibilities that exist amongst computer applications today.

\end{itemize}

The need for some discipline in how computing is done is not a new idea. In the early days of computing by machine, there was much `spaghetti programming' with the infamous ``goto'' statement, leading to the creation of programs that were difficult to understand and to maintain. This problem was largely solved by the introduction of `structured programming' (see, for example, \citet{jackson_1975}). Later, it became apparent that there could be more gains in the comprehensibility and maintainability of software via the introduction of `object-oriented' programming and design, modelling software on world-oriented objects and classes of object.

\subsection{Seamless integration of diverse kinds of knowledge and diverse aspects of intelligence}\label{seamless_integration_appendix}

In connection with the potential of the SP system as a UAI, an important third feature of the system, alongside its versatility in aspects of intelligence and its versatility in the representation of knowledge, is that {\em there is clear potential for the SP system to provide seamless integration of diverse kinds of knowledge and diverse aspects of intelligence, in any combination.} This is because diverse kinds of knowledge and diverse aspects of intelligence all flow from a single coherent and relatively simple source: SP-patterns within the SP-multiple-alignment framework.

In this respect, there is a sharp contrast between the SP system and the majority of other AI systems, which are either narrowly specialised for one or two functions or, if they aspire to be more general, are collections or kluges of different functions, with little or no integration.\footnote{Although Allen Newell called for the development of {\em Unified Theories of Cognition} \citep{newell_1992,newell_1990}, and researchers in `Artificial General Intelligence' are aiming for a similar kind of integration in AI, it appears that none of the resulting systems are fully integrated: ``We have not discovered any one algorithm or approach capable of yielding the emergence of [general intelligence].'' \citep[p.~1]{goertzel_2012}.}

This point is important because it appears that seamless integration of diverse kinds of knowledge and diverse aspects of intelligence, in any combination, are essential pre-requisites for human levels of fluidity, versatility and adaptability in intelligence.

\begin{figure}[!hbt]
\centering
\includegraphics[width=0.9\textwidth]{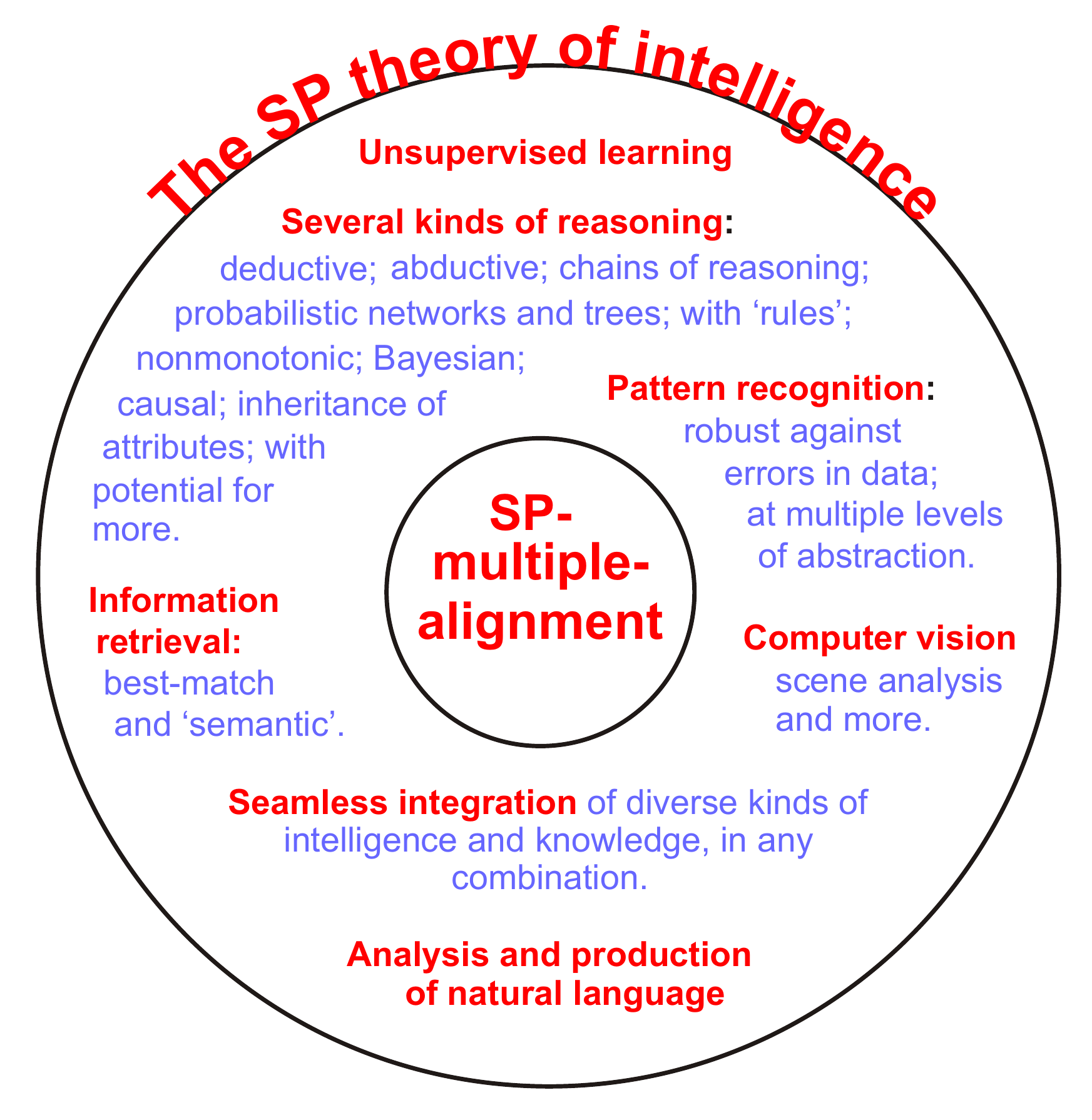}
\caption{A schematic representation of versatility and integration in the SP system, with SP-multiple-alignment centre stage.}
\label{versatility_integration_figure}
\end{figure}

Figure \ref{versatility_integration_figure} shows schematically how the SP system, with SP-multiple-alignment centre
stage, exhibits versatility and integration.

\subsection{Efficiency}\label{efficiency_appendix}

As noted near the beginning of Appendix \ref{uai_appendix}, the fifth suggested feature of a UAI is that it should in some sense be relatively `efficient' in its ability to represent diverse kinds of knowledge, to support diverse aspects of intelligence, and to provide for seamless integration of diverse kinds of knowledge and diverse aspects of intelligence, in any combination. This section expands on that idea.

It is anticipated that, when the SP system is more fully developed, it is likely to be more `efficient' than a CC, largely because it contains well-developed mechanisms for compression of information via the matching and unification of patterns (ICMUP), expressed via the powerful concept of SP-multiple-alignment. This provides the key to the SP system's versatility in the representation of diverse kinds of knowledge (Appendix \ref{versatility_in_rk_appendix}), its versatility in aspects of intelligence (Appendix \ref{versatility_in_intelligence_appendix}), and its potential for the seamless integration of diverse kinds of knowledge and diverse aspects of intelligence in any combination (Appendix \ref{seamless_integration_appendix}).

Although the computational `core' of a CC is likely to be smaller and simpler than in the SP Machine, the SP system has potential for relative advantages like these:

\begin{itemize}

    \item {\em More intelligence}. A CC is likely to fall short of the SP system in modelling the fluidity, versatility, and adaptability of human intelligence---unless the CC has been programmed with all the features of the SP system, in which case it would be an SP system and not a CC.

    \item {\em Economies in software}. Because of the pervasive influence of information compression in the SP system, its `software' is likely to be relatively compact. By contrast, the absence of well-developed mechanisms for ICMUP in the core of the CC is likely to mean the need for such mechanisms to be repeatedly recreated in different guises and in different applications. This can mean software with a bloating that more than offsets the small size of the central processor. See also Section \ref{overall_simplification_of_applications_section}.

    \item {\em Economies in data}. Unlike a CC, the SP system is designed to compress its data via unsupervised learning. This would normally mean that data for the SP Machine would, after compression, be substantially smaller than data for a CC.

    \item {\em Dramatic reductions in the variety of formats and formalisms}. As described in Appendix \ref{the_curse_of_variety_appendix}, an enormous variety of formats and formalisms is associated with conventional systems. The SP Machine has potential for dramatic simplifications in this area.

    \item {\em Efficiency in processing}. Although CCs, compared with human brains, are extraordinarily effective in such arithmetic tasks as adding up numbers or finding square roots, the advent of big data is creating demands that exceed the capabilities of the most powerful supercomputers \citep[p.~9]{kelly_hamm_2013}. But by exploiting statistical information that the SP system gathers as a by-product of how it works, there is potential in the system for substantial gains in the energy efficiency of its computations \citep[Sections VIII and IX]{sp_big_data}.

\end{itemize}

With regard to the second and third bullet points, all knowledge in the SP system reflects the world outside the system. This may include knowledge of entities and their interrelations---the kind of knowledge that would conventionally be called `data'---and knowledge of world-oriented processes or procedures---the kind of knowledge that might conventionally be called `software'.

All such knowledge is stored as SP-patterns without any formal distinctions amongst them. But in a CC, stored knowledge may be seen to comprise two components:

\begin{itemize}

    \item Knowledge of the system's environment, as in the SP Machine. This knowledge may be contained in external `databases' and also in `software'.

    \item Knowledge of processes or procedures, contained largely in `software', needed to overcome the deficiencies of the core model. This kind of knowledge, such as knowledge of how to search for matching SP-patterns, may be recreated many times in many different guises and in many different applications.

\end{itemize}

\section{The curse of variety in computing and what can be done about it}\label{the_curse_of_variety_appendix}

Wikipedia lists nearly 4,000 different `extensions' for computer files, representing a distinct type of file.\footnote{Details may be seen in ``List of filename extensions'', {\em Wikipedia}, \href{http://bit.ly/28LaT4v}{bit.ly/28LaT4v}, retrieved 2016-08-16.} A scan of the list suggests that most of these types of file are designed as input for this or that application. Each application is severely restricted in what kinds of file it can process---it is often only one---and incompatibilities are rife, even within one area of application such as word processing or the processing of images. And a program that will run on one operating system will typically not run on any other, so normally a separate version of each program is needed for each operating system, and, with some exceptions, each version needs its own kind of data file.

This kind of variety may also be found within individual files. In a Microsoft Word file, for example, there may be text in several different fonts and sizes, information generated by the ``track changes'' system, equations, WordArt, hyperlinks, bookmarks, cross-references, Clip Art, pre-defined shapes, SmartArt graphics, headers and footers, embedded Flash videos, images created by drawing tools, tables, and imported images in any of several formats including JPEG, PNG, Windows Metafile, and many more.

Excess variety is also alive and well amongst computer languages. Several hundred high-level programming languages are listed by Wikipedia, plus large numbers of assembly languages, machine languages, mark-up languages, style-sheet languages, query languages, modelling languages, and more.\footnote{There is more information in ``List of programming languages'', {\em Wikipedia}, \href{http://bit.ly/1GTW05W}{bit.ly/1GTW05W}, retrieved 2016-08-16; and also in ``Computer language'' and links from there, {\em Wikipedia}, \href{http://bit.ly/2aZ2kag}{bit.ly/2aZ2kag}, retrieved 2016-08-17.}

\subsection{Problems arising from excessive variety in computing}\label{problems_from_variety_appendix}

Excessive variety in computing is so familiar that we think of it as normal---part of the `wallpaper' of computing. But, although some may see that variety as evidence of vitality in computing, it is probably more accurate to see it as a symptom of a deep malaise in computing as it is today.

Much of this excessive variety is quite arbitrary, without any real justification, and the source of significant problems in computing such as:

\begin{itemize}

    \item {\em Bit rot}. The first of these, bit rot, is when software or data or both become unusable because technologies have moved on. Vint Cerf of Google has warned that the 21st century could become a second ``Dark Age'' because so much data is now kept in digital format, and that future generations would struggle to understand our society because technology is advancing so quickly that old files will be inaccessible. See, for example, ``Google's Vint Cerf warns of `digital Dark Age'\thinspace'', {\em BBC News}, 2015-02-13, \href{http://bbc.in/1D3pemp}{bbc.in/1D3pemp}.

    \item {\em Difficulties in extracting value from big data}. With big data---the humongous quantities of information that now flow from industry, commerce, science, and so on---excessive variety in formalisms and formats for knowledge and in how knowledge may be processed is one of several problems that make it difficult or impossible to obtain more than a small fraction of the value in those floods of data \citep{kelly_hamm_2013,national_research_council_2013}. Most kinds of processing---reasoning, pattern recognition, planning, and so on---will be more complex and less efficient than it needs to be \citep[Section III]{sp_big_data}. In particular, excess variety is likely to be a major handicap for data mining---the discovery of significant SP-patterns and structures in big data \citep[Section IV-B]{sp_big_data}.

    \item {\em Inefficiencies in the development of software}. Excessive variety in computing also means inefficiencies in the labour-intensive and correspondingly expensive process of developing software and the difficulty of reducing or eliminating bugs in software.

    \item {\em Safety and security}. And excess variety in computing means potentially serious consequences for such things as the safety of systems that depend on computers and software, and the security of computer systems. With regard to cybersecurity, Mike Walker, head of the Cyber Grand Challenge at DARPA, has said that it counts as a grand challenge because of, {\em inter alia}, the sheer complexity of modern software. A relevant news report is ``Can machines keep us safe from cyber-attack?'', {\em BBC News}, 2016-08-02, \href{http://bbc.in/2aLGwOu}{bbc.in/2aLGwOu}.

\end{itemize}

\subsection{A potential solution to the problem of excessive variety}\label{a_potential_solution_to_problems_of_variety_appendix}

The SP system provides a potential solution to the kinds of problems described in Appendix \ref{problems_from_variety_appendix}. It arises from the following three features of the SP system:

\begin{itemize}

    \item {\em Versatility of the SP system in the representation of knowledge}. The SP system has already-demonstrated versatility in the representation of diverse kinds of knowledge (Appendix \ref{versatility_in_rk_appendix}), with reasons to think that it may serve in the representation of {\em any} kind of knowledge (Appendix \ref{generality_in_rk_appendix}).

    \item {\em Versatility of the SP system in aspects of intelligence}. The SP system has already-demonstrated versatility in aspects of intelligence (Appendix \ref{versatility_in_intelligence_appendix}), with reasons to think that it provides a relatively firm foundation for the development of general, human-like artificial intelligence (Appendix \ref{generality_in_ai_appendix}).

    \item {\em Potential of the SP system to perform any kind of computable process or procedure}. As described in Appendix \ref{sp_programming_training_appendix}, the SP system has potential, via learning or programming, for any kind of computation that is not ruled out by problems with computational complexity.

\end{itemize}

An implication of the foregoing is that, instead of the great variety of kinds of input file for programs that prevails in computing today, we need only one: a type of computing file that contains SP-patterns, as described in Appendix \ref{overview_appendix}.

In a similar way, there is potential to replace all the many different computer languages with one language composed entirely of SP-patterns to be processed by the SP Machine.

A possible qualification to the idea that there might be only type of data file and one type of computer language is that, in both cases, users may wish to create sub-types of data file and sub-types of computer language by incorporating domain-specific knowledge in any given sub-type. For example, information about physics might be incorporated in a special-purpose language for use by physicists, and information about finance might be incorporated in a special-purpose language for that domain.


\begin{thebibliography}{32}
\providecommand{\natexlab}[1]{#1}
\providecommand{\url}[1]{\texttt{#1}}
\expandafter\ifx\csname urlstyle\endcsname\relax
  \providecommand{\doi}[1]{doi: #1}\else
  \providecommand{\doi}{doi: \begingroup \urlstyle{rm}\Url}\fi

\bibitem[Attneave(1954)]{attneave_1954}
F.~Attneave.
\newblock Some informational aspects of visual perception.
\newblock \emph{Psychological Review}, 61:\penalty0 183--193, 1954.

\bibitem[Barlow(1959)]{barlow_1959}
H.~B. Barlow.
\newblock Sensory mechanisms, the reduction of redundancy, and intelligence.
\newblock In {HMSO}, editor, \emph{The Mechanisation of Thought Processes},
  pages 535--559. Her Majesty's Stationery Office, London, 1959.

\bibitem[Barlow(1969)]{barlow_1969}
H.~B. Barlow.
\newblock Trigger features, adaptation and economy of impulses.
\newblock In K.~N. Leibovic, editor, \emph{Information Processes in the Nervous
  System}, pages 209--230. Springer, New York, 1969.

\bibitem[Birtwistle et~al.(1973)Birtwistle, Dahl, Myhrhaug, and
  Nygaard]{birtwistle_etal_1973}
G.~M. Birtwistle, O-J Dahl, B.~Myhrhaug, and K.~Nygaard.
\newblock \emph{Simula Begin}.
\newblock Studentlitteratur, Lund, 1973.

\bibitem[Church(1941)]{church_1941}
A.~Church.
\newblock \emph{The Calculi of Lamda-Conversion}, volume~6 of \emph{Annals of
  Mathematical Studies}.
\newblock Princeton University Press, Princeton, 1941.

\bibitem[Goertzel(2012)]{goertzel_2012}
B.~Goertzel.
\newblock Cogprime: an integrative architecture for embodied artificial general
  intelligence.
\newblock Technical report, The Open Cognition Project, 2012.
\newblock {PDF}: \href{http://bit.ly/1Zn0qfF}{bit.ly/1Zn0qfF}, 2012-10-02.

\bibitem[Jackson(1975)]{jackson_1975}
M.~A. Jackson.
\newblock \emph{Principles of Program Design}.
\newblock Academic Press, New York, 1975.

\bibitem[Kelly and Hamm(2013)]{kelly_hamm_2013}
J.~E. Kelly and S.~Hamm.
\newblock \emph{Smart machines: {IBM}'s {W}atson and the era of cognitive
  computing}.
\newblock Columbia University Press, New York, {K}indle edition, 2013.

\bibitem[Kirkpatrick(2017)]{kirkpatrick_2017}
J.~Kirkpatrick.
\newblock Overcoming catastrophic forgetting in neural networks.
\newblock \emph{Proceedings of the National Academy of Sciences of the United
  States of America}, 114\penalty0 (13):\penalty0 3521--3526, 2017.

\bibitem[Li and Vit{\'a}nyi(2014)]{li_vitanyi_2014}
M.~Li and P.~Vit{\'a}nyi.
\newblock \emph{An Introduction to Kolmogorov Complexity and Its Applications}.
\newblock Springer, New York, 3rd edition, 2014.

\bibitem[{National Research Council}(2013)]{national_research_council_2013}
{National Research Council}.
\newblock \emph{Frontiers in Massive Data Analysis}.
\newblock The National Academies Press, Washington DC, 2013.
\newblock {ISBN}-13: 978-0-309-28778-4. Online edition:
  \href{http://bit.ly/14A0eyo}{bit.ly/14A0eyo}.

\bibitem[Newell(1973)]{newell_1973}
A.~Newell.
\newblock You can't play 20 questions with nature and win: {P}rojective
  comments on the papers in this symposium.
\newblock In W.~G. Chase, editor, \emph{Visual Information Processing}, pages
  283--308. Academic Press, New York, 1973.

\bibitem[Newell(1990)]{newell_1990}
A.~Newell, editor.
\newblock \emph{Unified Theories of Cognition}.
\newblock Harvard University Press, Cambridge, Mass., 1990.

\bibitem[Newell(1992)]{newell_1992}
A.~Newell.
\newblock Pr{\'e}cis of {\em \uppercase{u}nified \uppercase{t}heories of
  \uppercase{c}ognition}.
\newblock \emph{Behavioural and Brain Sciences}, 15\penalty0 (3):\penalty0
  425--437, 1992.

\bibitem[Palade and Wolff(2017)]{devt_sp_machine}
V.~Palade and J.~G. Wolff.
\newblock Development of a new machine for artificial intelligence.
\newblock Technical report, CognitionResearch.org, 2017.
\newblock Submitted for publication.
  \href{http://bit.ly/2tWb88M}{bit.ly/2tWb88M}, arXiv:1707.0061 [cs.AI].

\bibitem[Post(1943)]{post_1943}
E.~L. Post.
\newblock Formal reductions of the general combinatorial decision problem.
\newblock \emph{American Journal of Mathematics}, 65:\penalty0 197--268, 1943.

\bibitem[Solomonoff(1964)]{solomonoff_1964}
R.~J. Solomonoff.
\newblock A formal theory of inductive inference. {P}arts {I} and {II}.
\newblock \emph{Information and Control}, 7:\penalty0 1--22 and 224--254, 1964.

\bibitem[Solomonoff(1997)]{solomonoff_1997}
R.~J. Solomonoff.
\newblock The discovery of algorithmic probability.
\newblock \emph{Journal of Computer and System Sciences}, 55\penalty0
  (1):\penalty0 73--88, 1997.

\bibitem[Turing(1936)]{turing_1936}
A.~M. Turing.
\newblock On computable numbers with an application to the
  {E}ntscheidungsproblem.
\newblock \emph{Proceedings of the London Mathematical Society}, 42:\penalty0
  230--265 and 544--546, 1936.

\bibitem[Wolff(1988)]{wolff_1988}
J.~G. Wolff.
\newblock Learning syntax and meanings through optimization and distributional
  analysis.
\newblock In Y.~Levy, I.~M. Schlesinger, and M.~D.~S. Braine, editors,
  \emph{Categories and Processes in Language Acquisition}, pages 179--215.
  Lawrence Erlbaum, Hillsdale, NJ, 1988.
\newblock \href{http://bit.ly/ZIGjyc}{bit.ly/ZIGjyc}.

\bibitem[Wolff(2006{\natexlab{a}})]{wolff_2006}
J.~G. Wolff.
\newblock \emph{Unifying Computing and Cognition: the {SP} Theory and Its
  Applications}.
\newblock CognitionResearch.org, Menai Bridge, 2006{\natexlab{a}}.
\newblock {ISBN}s: 0-9550726-0-3 (ebook edition), 0-9550726-1-1 (print
  edition). Distributors, including Amazon.com, are detailed on
  \href{http://bit.ly/WmB1rs}{bit.ly/WmB1rs}.

\bibitem[Wolff(2006{\natexlab{b}})]{wolff_medical_diagnosis}
J.~G. Wolff.
\newblock Medical diagnosis as pattern recognition in a framework of
  information compression by multiple alignment, unification and search.
\newblock \emph{Decision Support Systems}, 42:\penalty0 608--625,
  2006{\natexlab{b}}.
\newblock \doi{10.1016/j.dss.2005.02.005}.
\newblock \href{http://bit.ly/1F366o7}{bit.ly/1F366o7}, arXiv:1409.8053
  [cs.AI].

\bibitem[Wolff(2007)]{wolff_sp_intelligent_database}
J.~G. Wolff.
\newblock Towards an intelligent database system founded on the {SP} theory of
  computing and cognition.
\newblock \emph{Data \& Knowledge Engineering}, 60:\penalty0 596--624, 2007.
\newblock \doi{10.1016/j.datak.2006.04.003}.
\newblock \href{http://bit.ly/1CUldR6}{bit.ly/1CUldR6}, arXiv:cs/0311031
  [cs.DB].

\bibitem[Wolff(2013)]{sp_extended_overview}
J.~G. Wolff.
\newblock The {SP} theory of intelligence: an overview.
\newblock \emph{Information}, 4\penalty0 (3):\penalty0 283--341, 2013.
\newblock \doi{10.3390/info4030283}.
\newblock \href{http://bit.ly/1NOMJ6l}{bit.ly/1NOMJ6l}, arXiv:1306.3888
  [cs.AI].

\bibitem[Wolff(2014{\natexlab{a}})]{sp_autonomous_robots}
J.~G. Wolff.
\newblock Autonomous robots and the {SP} theory of intelligence.
\newblock \emph{IEEE Access}, 2:\penalty0 1629--1651, 2014{\natexlab{a}}.
\newblock \doi{10.1109/ACCESS.2014.2382753}.
\newblock \href{http://bit.ly/18DxU5K}{bit.ly/18DxU5K}, arXiv:1409.8027
  [cs.AI].

\bibitem[Wolff(2014{\natexlab{b}})]{sp_benefits_apps}
J.~G. Wolff.
\newblock The {SP} theory of intelligence: benefits and applications.
\newblock \emph{Information}, 5\penalty0 (1):\penalty0 1--27,
  2014{\natexlab{b}}.
\newblock \doi{10.3390/info5010001}.
\newblock \href{http://bit.ly/1FRYwew}{bit.ly/1FRYwew}, arXiv:1307.0845
  [cs.AI].

\bibitem[Wolff(2014{\natexlab{c}})]{sp_big_data}
J.~G. Wolff.
\newblock Big data and the {SP} theory of intelligence.
\newblock \emph{IEEE Access}, 2:\penalty0 301--315, 2014{\natexlab{c}}.
\newblock \doi{10.1109/ACCESS.2014.2315297}.
\newblock \href{http://bit.ly/2qfSR3G}{bit.ly/2qfSR3G}, arXiv:1306.3890
  [cs.DB]. This paper, with minor revisions, is reproduced in Fei Hu (Ed.),
  {\em Big Data: Storage, Sharing, and Security}, Taylor \& Francis LLC, CRC
  Press, 2016, Chapter 6, pp.~143--170.

\bibitem[Wolff(2014{\natexlab{d}})]{sp_vision}
J.~G. Wolff.
\newblock Application of the {SP} theory of intelligence to the understanding
  of natural vision and the development of computer vision.
\newblock \emph{SpringerPlus}, 3\penalty0 (1):\penalty0 552--570,
  2014{\natexlab{d}}.
\newblock \doi{10.1186/2193-1801-3-552}.
\newblock \href{http://bit.ly/2oIpZB6}{bit.ly/2oIpZB6}, arXiv:1303.2071
  [cs.CV].

\bibitem[Wolff(2016{\natexlab{a}})]{sp_alternatives}
J.~G. Wolff.
\newblock The {SP} theory of intelligence: its distinctive features and
  advantages.
\newblock \emph{IEEE Access}, 4:\penalty0 216--246, 2016{\natexlab{a}}.
\newblock \doi{10.1109/ACCESS.2015.2513822}.
\newblock \href{http://bit.ly/2qgq5QF}{bit.ly/2qgq5QF}, arXiv:1508.04087
  [cs.AI].

\bibitem[Wolff(2016{\natexlab{b}})]{sp_csrk}
J.~G. Wolff.
\newblock Commonsense reasoning, commonsense knowledge, and the {SP} theory of
  intelligence.
\newblock Technical report, CognitionResearch.org, 2016{\natexlab{b}}.
\newblock Submitted for publication.
  \href{http://bit.ly/2eBoE9E}{bit.ly/2eBoE9E}, arXiv:1609.07772 [cs.AI].

\bibitem[Wolff(2017)]{sp_compression}
J.~G. Wolff.
\newblock Information compression via the matching and unification of patterns
  as a unifying principle in human learning, perception, and cognition.
\newblock Technical report, CognitionResearch.org, 2017.
\newblock Submitted for publication.
  \href{http://bit.ly/2ruLnrV}{bit.ly/2ruLnrV}, viXra:1707.0161v2,
  hal-01624595, v1.

\bibitem[Wolff(2018)]{sp_intro_2018}
J.~G. Wolff.
\newblock Introduction to the {SP} theory of intelligence.
\newblock Technical report, CognitionResearch.org, 2018.
\newblock arXiv:1802.09924 [cs.AI],
  \href{http://bit.ly/2ELq0Jq}{bit.ly/2ELq0Jq}.

\end{thebibliography}

\end{document}